\documentclass[ALICE,manyauthors]{cernphprep}
\usepackage[comma,square,numbers,sort&compress]{natbib}
\usepackage[dvipsnames]{xcolor}

\usepackage{lineno}
\modulolinenumbers[5]
\usepackage{xspace}
\usepackage{hyperref}
\usepackage[T1]{fontenc}
\usepackage{orcidlink}

\newcommand{\pp}           {pp\xspace}

\newcommand{\PbPb}         {\mbox{Pb--Pb}\xspace}

\newcommand{\snn}          {\ensuremath{\sqrt{s_{\mathrm{NN}}}}\xspace}

\newcommand{\nineH}        {$\sqrt{s}~=~0.9$~Te\kern-.1emV\xspace}
\newcommand{\seven}        {$\sqrt{s}~=~7$~Te\kern-.1emV\xspace}
\newcommand{\twoH}         {$\sqrt{s}~=~0.2$~Te\kern-.1emV\xspace}
\newcommand{\twosevensix}  {$\sqrt{s}~=~2.76$~Te\kern-.1emV\xspace}
\newcommand{\five}         {$\sqrt{s}~=~5.02$~Te\kern-.1emV\xspace}
\newcommand{\twosevensixnn}{$\sqrt{s_{\mathrm{NN}}}~=~2.76$~Te\kern-.1emV\xspace}
\newcommand{\fivenn}       {$\sqrt{s_{\mathrm{NN}}}~=~5.02$~Te\kern-.1emV\xspace}

\newcommand{\GeVc}         {Ge\kern-.1emV/$c$\xspace}
\newcommand{\MeVc}         {Me\kern-.1emV/$c$\xspace}
\newcommand{\TeV}          {Te\kern-.1emV\xspace}
\newcommand{\GeV}          {Ge\kern-.1emV\xspace}
\newcommand{\MeV}          {Me\kern-.1emV\xspace}
\newcommand{\GeVmass}      {Ge\kern-.2emV/$c^2$\xspace}
\newcommand{\MeVmass}      {Me\kern-.2emV/$c^2$\xspace}

\begin{document}

\begin{titlepage}
\PHyear{2022}       
\PHnumber{072}      
\PHdate{29 March}  

\title{ALICE luminosity determination for \PbPb collisions at~$\mathbf{\sqrt{\emph{s}_{\mathrm{NN}}}~=~5.02}$~Te\kern-.1emV}
\ShortTitle{Luminosity determination for \PbPb collisions at \fivenn}   

\Collaboration{ALICE Collaboration\thanks{See Appendix~\ref{app:collab} for the list of collaboration members}}
\ShortAuthor{ALICE Collaboration} 

\begin{abstract}
Luminosity determination within the  ALICE experiment is based on the measurement, in van der Meer scans, of the cross sections for visible processes involving one or more detectors (visible cross sections). In 2015 and 2018, the Large Hadron Collider provided \PbPb collisions at a centre-of-mass energy per nucleon pair of \fivenn. Two visible cross sections, associated with particle detection in the Zero Degree Calorimeter (ZDC) and in the V0 detector, were measured in a van der Meer scan.
This article describes the experimental set-up and the analysis procedure, and presents the measurement results. The analysis involves a comprehensive study of beam-related effects and an improved fitting procedure, compared to previous ALICE studies, for the extraction of the visible cross section. The resulting uncertainty of both the ZDC-based and the V0-based luminosity measurement for the full sample is 2.5\%. The inelastic cross section for hadronic interactions in \PbPb collisions at \fivenn, obtained by efficiency correction of the V0-based visible cross section, was measured to be 7.67~$\pm$~0.25~b, in agreement with predictions using the Glauber model.

\end{abstract}
\end{titlepage}

\setcounter{page}{2} 


\section{Introduction} 

Cross section measurements at colliders require precise luminosity determination. The rate $R$ of a process can be expressed as
\begin{equation}
    R =  L \sigma,
\end{equation}
where $L$ is the luminosity and $\sigma$ is the process cross section.

In a bunched circular collider, such as the Large Hadron Collider~\cite{Evans:2008zzb} (LHC), the particles circulate in packets (bunches) of finite length, defined by the radio-frequency structure of the accelerator. For two contra-rotating bunches colliding with a null crossing angle, the luminosity can be expressed as
\begin{equation}\label{eq:lumiDef}
    L = \nu_{\rm{rev}}N_1N_2\int{f_1(x,y)f_2(x,y){\rm{d}}x{\rm{d}}y},
\end{equation}
where $\nu_{\rm{rev}}$ is the accelerator revolution frequency, $N_1$ and $N_2$ are the bunch intensities, defined as the number of particles in a bunch, $f_1$ and $f_2$ are the probability density distributions of particles in the transverse ($x$,$y$) plane for the two bunches (where $x$ is the horizontal direction and $y$ is the vertical direction), assumed to be independent of the longitudinal coordinate $z$. A detailed discussion of the concept and definition of luminosity can be found in~\cite{Herr:2003em}.

Assuming factorisation of the density distributions in the two transverse directions, such that
\begin{equation}
    f_1(x,y) = f_{1x}(x)f_{1y}(y),\;\; f_2(x,y) = f_{2x}(x)f_{2y}(y),
\end{equation}
one can write
\begin{equation}
    L = \frac{\nu_{\rm{rev}}N_1N_2}{h_xh_y},
    \label{eq:lumi}
\end{equation}
where
\begin{equation}\label{eq:effWitdh}
    h_x = \frac{1}{\int{f_{1x}(x)f_{2x}(x){\rm{d}}x}}\;\;{\rm{and}}\;\;h_y = \frac{1}{\int{f_{1y}(y)f_{2y}(y){\rm{d}}y}}
\end{equation}
are the effective widths of the beam overlap region.

The van der Meer (vdM) scan~\cite{vdM} is the most common technique employed for luminosity determination at colliders, see, e.g. ~\cite{Grafstrom:2015foa} for a review, and~\cite{atlasLumi0,atlasLumi1,atlasLumi2,CMS:2021xjt,lhcbLumi, lhcbLumi2,Abelev:2012sea,aliVdMpPb} for measurements performed at the LHC.
In vdM scans, the two beams are moved in the transverse plane, in discrete steps. The rate $R_{\rm{vis}}$($\Delta x$, $\Delta y$) of a reference (visible) process is measured as a function of the transverse beam separations ($\Delta x$, $\Delta y$), defined as the distance between the centroids of the beam bunches. The $x$ and $y$ scans are usually performed separately, the beams being head-on (i.e. colliding with zero separation) in the non-scanned direction. In this case, the effective widths $h_{x0}$ and $h_{y0}$ for head-on collisions 
can be determined as
\begin{equation}
    h_{x0} = \frac{\int{R_{\rm{vis}}(\Delta x,0){\rm{d}}\Delta x}}{R_{\rm{vis}}(0,0)},\;\;h_{y0} = \frac{\int{R_{\rm{vis}}(0, \Delta y){\rm{d}}\Delta y}}{R_{\rm{vis}}(0,0)}
    \label{eq:vdmWidth}
\end{equation}
(see ~\cite{vdM} for a derivation).

When the beams collide with a non-zero crossing angle, Eqs.~\ref{eq:lumiDef} and~\ref{eq:effWitdh} need to be modified~\cite{Herr:2003em}, but it can be shown~\cite{balagura} that the vdM scan technique still allows a precise luminosity determination, and, in particular, that Eqs.~\ref{eq:lumi} and~\ref{eq:vdmWidth} still hold.

The main output of vdM scans is a measurement of the cross section $\sigma_{\rm{vis}}$ of the visible process, which can be determined as
\begin{equation}
    \sigma_{\rm{vis}} = 
    \frac{R_{\rm{vis}}(0,0)\,h_{x0} h_{y0}}{\nu_{\rm{rev}}N_1N_2}
    \label{eq:crSec}
\end{equation}
and used for the measurement of luminosity during physics data-taking:
\begin{equation}
    L = \frac{R_{\rm{vis}}}{\sigma_{\rm{vis}}}.
    \label{eq:lumiFromRvis}
\end{equation}

The standard vdM scans are typically coupled with a length-scale calibration scan, whose aim is to determine the global conversion factor from the nominal beam displacement 
to the actual one. In these scans, the two beams are kept at constant separation and moved in consecutive steps in the same direction, and the position of the interaction vertex is measured, using the tracking detectors, as a function of the nominal beam position.

The vdM formalism assumes complete factorisation of the beam profiles in the two transverse directions, such that the beam overlap region is fully described by the product $h_x h_y$. Previous studies performed by ALICE~\cite{aliVdMpPb,aliceLumi13TeV,aliceLumi8TeV,ALICE-PUBLIC-2016-005,
ALICE-PUBLIC-2018-014,
ALICE-PUBLIC-2018-002} and other LHC experiments~\cite{lhcbLumi2,atlasLumi1,CMS:2021xjt, atlasLumi2,cmsLumi1} have shown that the actual LHC bunch shapes can violate the factorisation assumption. The size of the effect was found to vary from scan to scan and demanded corrections ranging from the per mil to the percent level. Non-factorisation effects can be studied and quantified by measuring the luminous region parameters, via the distribution of interaction vertices, as a function of the beam separation.

During Run 2,  the LHC provided, in 2015 and 2018, lead--lead (\PbPb) collisions at a centre-of-mass energy per nucleon pair \fivenn. 
For collisions of lead ions, the visible cross section $\sigma_{\rm{vis}}$ seen by a detector (or set of detectors) with a given trigger condition has, in general, two components, one electromagnetic and one hadronic: $\sigma_{\rm{vis}}$~$=$~$\epsilon_{\rm{EM}} \sigma_{\rm{EM}} + \epsilon_{\rm{had}} \sigma_{\rm{had}}$, where $\sigma_{\rm{EM}}$ and $\sigma_{\rm{had}}$ are the electromagnetic  and hadronic inelastic cross sections and $\epsilon_{\rm{EM}}$ and $\epsilon_{\rm{had}}$ are the fractions of electromagnetic and hadronic inelastic events that satisfy the trigger condition.

The ALICE luminosity determination for the Run 2 Pb--Pb data samples is based on a vdM scan session 
that took place on November 29, 2018, during the LHC fill~\footnote{A fill is a time interval with continued presence of beam in the accelerator; it starts with the injection and ends with the beam dump.} labelled with the number 7483. 
The visible cross sections for two independent reference processes were measured in this scan session, and used for the indirect luminosity determination of the 2015 and 2018 samples, according to Eq.~\ref{eq:lumiFromRvis}. Note that this procedure does not require a knowledge of $\epsilon_{\rm{had}}$ or $\epsilon_{\rm{EM}}$.

This document is organised as follows. Section~\ref{set-up} describes the detectors used for the measurement, along with the relevant machine parameters and the procedure adopted for the scan. Section~\ref{sec:results} summarises the analysis procedure and presents the results and uncertainties for the visible cross section and luminosity measurement, and for the inelastic hadronic cross section for Pb--Pb collisions at \fivenn. The hadronic cross section was determined by combining one of the measured visible cross sections and a data-driven estimate of the corresponding hadronic efficiency $\epsilon_{\rm{had}}$.   Finally, Sec.~\ref{sec:Conclusions} presents a brief summary of the work.

\section{Experimental set-up \label{set-up}}

In the vdM scan, the cross section was measured for two reference processes, one triggered upon by the Zero Degree Calorimeter (ZDC), the other by the V0 detector. A detailed description of these detectors is given in~\cite{ALICE_jinst}, and their performance is discussed in~\cite{aliPerf,vzeroPerf}. The ZDC system features two neutron calorimeters (ZNA, ZNC), located on opposite sides of the ALICE interaction point (IP2), each one at a distance of 112.5~m along the beam axis from IP2, covering the pseudorapidity ($\eta$) range ${\left|\eta\right|>8.8}$. It is completed by two proton calorimeters and two electromagnetic calorimeters, not used for this measurement. The V0 detector consists of two hodoscopes, with 32 scintillator tiles each, located on opposite sides of the interaction region, at distances of 340~cm (V0A) and 90~cm (V0C) along the beam axis from IP2, covering the pseudorapidity ranges ${2.8<\eta< 5.1}$ and $-3.7<\eta<-1.7$, respectively. Note that the LHC beam 1 (2) travels clockwise (anticlockwise) from side A (C) to side C (A). 

The ZDC-based visible cross section is defined by a trigger condition, called ZED in the following, which requires a signal in at least one of the two neutron calorimeters, corresponding to an energy deposition larger than~$\sim$~1~TeV. Such a threshold is about three standard deviations below the expected signal from a 2.51~TeV neutron. Neutrons are emitted from the fragmentation/evaporation of Pb ions in electromagnetic dissociation events with (single- or double-side) neutron emission, or in hadronic events~\cite{BaltzEtal1996,ALICE:2012aa,Pshenichnov2001,Pshenichnov2011,Broz:2019kpl}. The trigger condition for the V0-based visible cross section, called V0M in the following, requires the sum of the signal amplitudes from all the V0 scintillators to be above a chosen threshold; during the 2018 \PbPb data taking, the threshold was such that the $\sim$~50\% most central hadronic events were selected, and all electromagnetic events were rejected due to their relatively low particle multiplicity in the V0 acceptance.

The analysis procedure uses, for the length-scale calibration and non-factorisation corrections, the parameters of the luminous region measured via the distribution of interaction vertices, determined with the ALICE Inner Tracking System~\cite{aliITS} (ITS).

During the vdM scan session, each Pb beam consisted of 648 bunches, and 619 bunch pairs were colliding at IP2. The minimum spacing between two consecutive bunches in each beam was 100~ns.
The $\beta^*$ value~\footnote{The $\beta(z)$ function describes the single-particle motion and determines the variation of the beam envelope as a function of the coordinate along the beam orbit ($z$). The transverse size of the beam at a given position along the beam trajectory is proportional to the square root of $\beta$. The notation $\beta^*$ denotes the value of the $\beta$ function at the interaction point.} at IP2 was 0.5~m.
The nominal half vertical crossing angle of the two beams at IP2 was about~$-60~\mu$rad, the minus sign indicating that the two beams exited the crossing region with negative $y$ coordinate with respect to the beam axis~\footnote{ALICE uses a Cartesian system whose origin is at the LHC Interaction Point 2 (IP2). The $z$ axis is parallel to the mean beam direction at IP2 and points along the LHC Beam 2 (i.e. LHC anticlockwise). The $y$ axis points upwards while the $x$ axis is perpendicular to the $y$ and $z$ axes, forming a right-handed orthogonal system.
}. The current in the ALICE solenoid (dipole) was 30~kA (6~kA),  corresponding to a field strength of 0.5~T (0.7~T).  

Two pairs of horizontal and vertical scans were performed, to obtain two statistically independent cross section measurements per bunch pair. In each horizontal (vertical) scan, the nominal beam separation $\Delta x$ ($\Delta y) $ was varied in 25 equal steps~\footnote{See Appendix~\ref{app:function} for details} from $-97.3$~$\mu$m to $+97.3$~$\mu$m. A separation of 100~$\mu$m corresponds to about six times the root mean square of the transverse beam profile.  During each step, the beams were maintained in position for 28~s, and the ZED and V0M trigger counts were integrated in 14 time bins of 2~s each. The counts were measured separately for each colliding bunch pair.   In order to provide additional input for non-factorisation studies, two diagonal scans were performed, where the beam separation was varied simultaneously in the two transverse directions. Finally, a set of length-scale calibration scans was performed.

The bunch intensities were of the order of (7--10)$\times$10$^{7}$ Pb ions per bunch. 
The bunch-intensity measurement was provided by the LHC instrumentation~\cite{beam_current}: a direct current transformer (DCCT), measuring the total beam intensity, and a fast beam current transformer (fBCT), measuring the relative bunch intensities. For the relative bunch intensities, data from a second device, the ATLAS beam pick-up system (BPTX~\cite{atlas_bptx}) was also used. 
The accelerator orbit is nominally divided in 3564 slots of 25~ns each. Given the radio-frequency configuration of the LHC, each slot is divided in ten buckets of 2.5~ns each. In nominally filled slots, the so-called main bunch is captured in the central bucket of the slot. Following the convention established in~\cite{bcnwg_long}, the charge circulating outside of the nominally filled slots is referred to as ghost charge;  the charge circulating within a nominally filled slot but not captured in the central bucket is referred to as satellite charge. The ghost and satellite charges do not contribute to the luminosity at the nominal interaction point. Hence, they must be subtracted from the total beam intensity.  
A measurement of the ghost-charge fraction was provided independently by the LHCb collaboration, via the rate of beam--gas collisions~\footnote{A definition of beam--gas collision is provided in Section~\ref{sec:results}.} occurring in nominally empty bunch slots, as described in~\cite{lhcbLumi2}, and by the LHC Longitudinal Density Monitor (LDM), which measures synchrotron radiation photons emitted by the beams~\cite{ldm_note}.  
The LDM also provides a measurement of the satellite-charge fraction.
For the vdM scan under analysis, the measured ghost-charge fraction was about 4\% (3\%) for beam 1 (beam 2) and the bunch-averaged satellite-charge fraction was about 3\% for both beams, resulting in a total correction to the bunch intensity product (hence to the cross section) of about~13\%. Satellite bunches in a beam may interact with main bunches in the other beam. These events must be identified and subtracted from the measured visible process rates, as will be described in Section~\ref{sec:results}.

\section{Analysis and results} \label{sec:results} 

\subsection{Visible cross section determination}

In previous studies dedicated to the luminosity determination in \pp~\cite{aliceLumi13TeV,aliceLumi8TeV,ALICE-PUBLIC-2016-005,
ALICE-PUBLIC-2018-014}, p--Pb~\cite{aliVdMpPb,ALICE-PUBLIC-2018-002}, and \hbox{Pb--Pb}~\cite{aliPerf} collisions in the ALICE experiment, the trigger rates were measured as a function of the beam separation and corrected for background and pile-up effects. A $\chi^2$-based fit of the scan curves (separately for the $x$ and $y$ scans) yielded a measurement of $R(0,0)$, $h_{x0}$, and $h_{y0}$, which could be inserted directly into Eq.~\ref{eq:crSec} to determine $\sigma_{\rm{vis}}$. In comparison, e.g. to the studies performed for \pp collisions, the present analysis deals with a collision rate per colliding bunch pair lower by about one order of magnitude for ZED and three orders of magnitude for V0M. This demands a different approach, designed to obtain a better treatment of statistical uncertainties at very small numbers of trigger counts. For each colliding bunch pair, the number of triggered events $t_i$ and the number of sampled LHC orbits $n_i$ during time bin $i$ are used as inputs for a binomial likelihood fit:
\begin{equation}\label{eq:likelihood}
\ln{\mathcal{L}}={\displaystyle \sum_{i}}\left[t_i\ln P_i+\left(n_{i}-t_i\right)\ln{(1-P_i)}\right]
\end{equation}
 where $P_i$ is the probability of having a trigger in a bunch crossing, related to the mean number of triggers per bunch crossing  $\mu_i$ by Poissonian statistics, $P_i=1-e^{-\mu_{i}}$.
The quantity $\mu_i$ is modelled by the fit function, according to the relations
\begin{equation}\label{eq:mu}
    \mu_i = \frac{R_{\rm{vis}}(\Delta x_i, \Delta y_i)}{\nu_{\rm{rev}}} +p_{\rm{s},\it{i}} +\tilde{p}_1 N_{1,i} + \tilde{p}_2 N_{2,i}+p_0
\end{equation}

and 

\begin{equation} \label{eq:R_vis}
    R_{\rm{vis}}(\Delta x_i, \Delta y_i) = \nu_{\rm{rev}}N_{1,i}N_{2,i}\frac{\sigma_{\rm{vis}}}{h_{x0} h_{y0}}f(\Delta x_i)g(\Delta y_i),
\end{equation}
where: $N_{1,i}$ and $N_{2,i}$ are the intensities of the two colliding bunches; $\Delta x_i$ and $\Delta y_i$ are the beam separations, corrected for beam--beam deflection~\cite{kicknote,balaguraBeamBeam} and orbit drifts~\cite{bpm,yasp}, $f$ and $g$ parametrise the luminosity dependence on $\Delta x_i$ and $\Delta y_i$, respectively; $h_{x0}$ and $h_{y0}$ are the integrals of $f$ and $g$, respectively, 
divided by their peak values, consistently with Eq.~\ref{eq:vdmWidth}; $p_{\rm{s},\it{i}}$ is the separation-dependent probability that the trigger is fired by a collision between one of the two colliding bunches and a satellite bunch in the other beam, or by the collision of two satellites; $\tilde{p}_{1}$ ($\tilde{p}_{2}$) is the probability that the trigger is fired by a collision of a bunch of beam~1 (beam~2) with residual gas in the beam pipe (beam--gas collision), normalised by the bunch intensity; $p_0$ is the probability that the trigger is fired in the absence of beams (detector noise).

The functions $f$($\Delta x$) and $g$($\Delta y$) were chosen to have a Gaussian core with mean value and standard deviation as the only free parameters, the normalisation being constrained by Eq.~\ref{eq:crSec}.
In order to improve the description of data at large separation, the Gaussian function is modified at absolute separations larger than a certain threshold. For each scan step beyond the threshold, an independent offset is added to $\Delta x_i$ or $\Delta y_i$ in the definition of the fitting function, so that there is one additional fit parameter for each of these steps. The threshold is chosen, independently for each colliding bunch pair, as the minimum value allowing one to obtain $\chi^2/\textit{ndf}~\sim~1$; depending on the considered colliding bunch pair and scan, it is located 1.3--2.5 standard deviations away from the peak, and the total number of parameters needed to describe the tails varies between 7 and 13. The function is constrained to be symmetric around the peak by using the same tail parameter for scan steps at opposite nominal separation. A formal definition of the fitting functions $f$($\Delta x$) and $g$($\Delta y$) is provided in Appendix~\ref{app:function}.

The parameters $p_0$, $\tilde{p}_1$ and $\tilde{p}_2$ were estimated by means of an independent fit to the trigger rates in non-colliding and empty bunch slots. Empty bunch slots located immediately after colliding bunch slots were excluded from the fit, because such bunch slots are affected by background from late spurious pulses (after-pulses) and would provide an overestimated measurement of the detector noise. Owing to the minimum spacing of 100~ns between colliding bunches, the contribution from a previous collision to the trigger counts in colliding bunch slots was found to be negligible for both ZED and V0M signals.   Because of the large ZDC distance from IP2, the background induced on ZNA (ZNC) by beam--gas collisions of a bunch of beam 1 (2) happening upstream of the calorimeter results in a signal that is 31 bunch slots ($\sim$~750~ns) earlier with respect to nominal beam--beam collisions of that bunch. During the vdM scan, the distribution of Pb-ion bunches along the LHC orbit was such that this background contribution shows up only in nominally empty bunch slots, with no effect on the colliding slots. Therefore, for the ZED analysis, this subset of the empty bunch slots was excluded from the background fit.

\begin{figure}
\includegraphics[width=0.5\columnwidth]{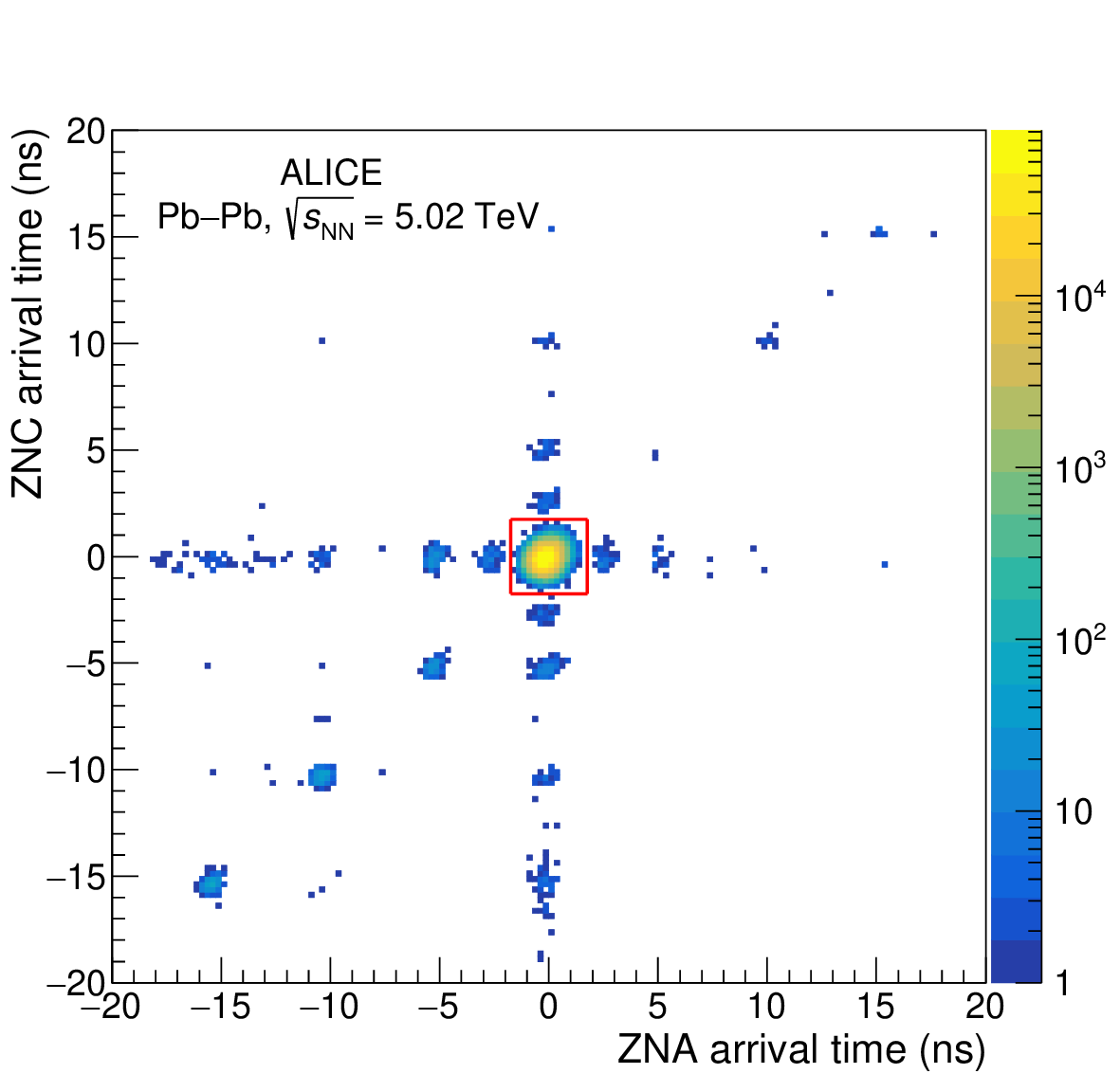}
\includegraphics[width=0.5\columnwidth]{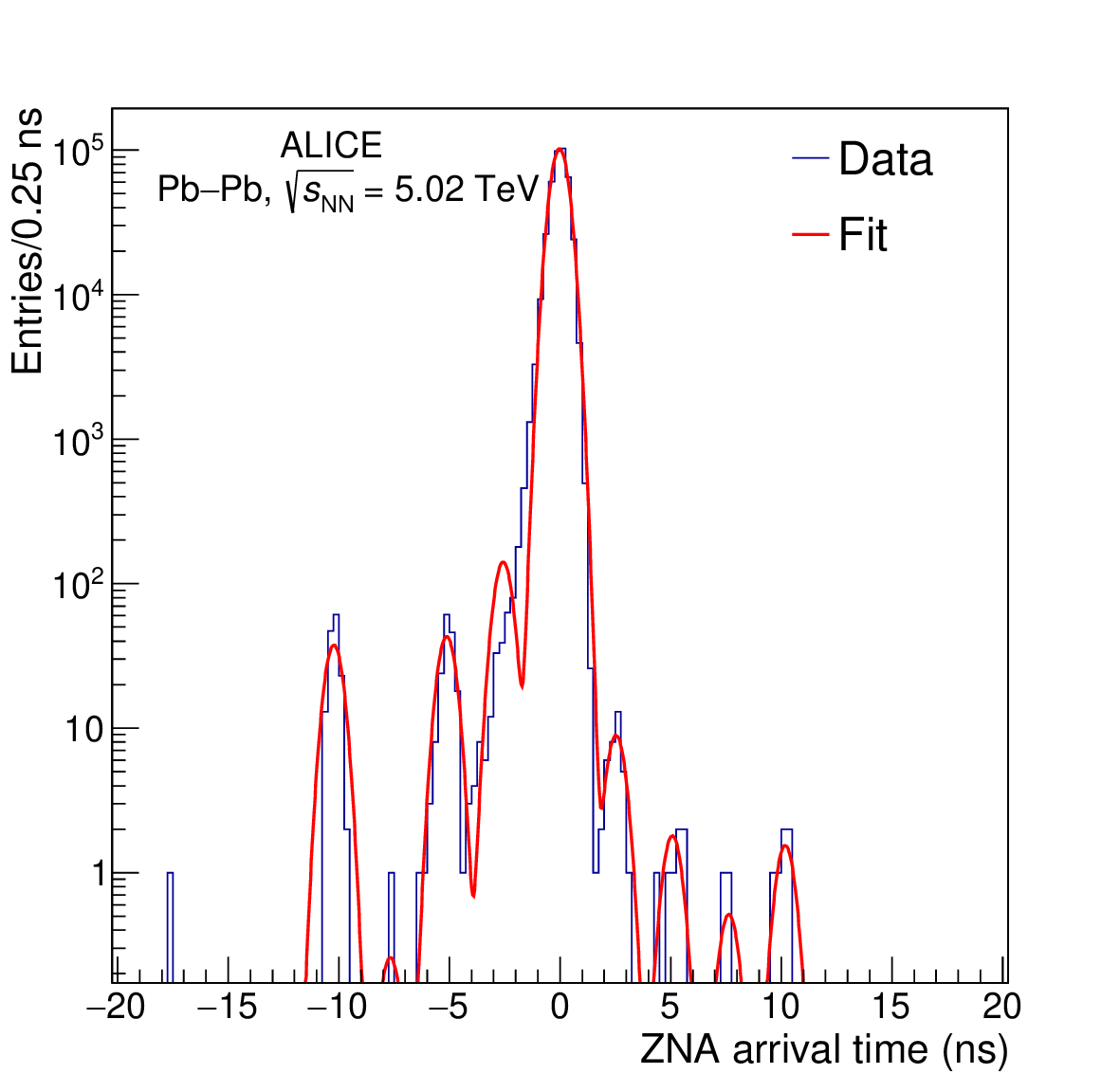}
\caption{\label{fig:zdc_tdc}(Colour online) Left: correlation
between the arrival times, relative to main--main collisions, of signals in ZNA and ZNC  for events triggered by V0M. The square box depicts the satellite-collision rejection cut discussed in the text. Events outside the box are from main--satellite collisions (horizontal and vertical bands) and from satellite--satellite collisions (diagonal band).
Right: distribution of the arrival time, relative to main-main collisions, in one of the two neutron calorimeters for ZED-triggered single-side neutron events. The superimposed curve shows a fit with a sum of Gaussian distributions. Both figures are for head-on collisions ($\Delta x=\Delta y=0$).
}
\end{figure}

The separation-dependent contribution from main--satellite collisions $p_{\rm{s},\it{i}}$ was evaluated 
via the signal arrival-time spectra in ZNA and ZNC. The procedure is different for ZED and V0M due to the different selectivity of the two trigger classes. All events triggered by V0M are hadronic and have signals in both ZNA and ZNC. The two-dimensional distribution of arrival times in the two calorimeters for these events is shown in the left panel of Fig.~\ref{fig:zdc_tdc}. The satellite events are tagged by means of a square cut around the main--main collision peak position, located at (0, 0). Conversely, the ZED trigger has a large contribution from electromagnetic events with single-side neutron emission, so that most of the events have a signal only in one calorimeter. For this sub-sample of ZED-triggered events the estimation of the satellite contamination is based on the one-dimensional arrival time distributions in each of the ZNs, and the fraction of satellite collisions is obtained via a fit of the time distribution to a sum of Gaussian functions, with peak positions fixed to the values expected from the LHC radio-frequency structure (right panel of Fig.~\ref{fig:zdc_tdc}).
The signal from a neutron emitted in a main--satellite collision has the same arrival time as that from a main--main collision if the neutron is emitted by an ion in the main bunch, while it is early or late if the neutron is emitted by an ion in the satellite bunch. Therefore, only half of the neutrons emitted in single-side events from main--satellite collisions are identified as such. Hence, a correction factor of two was applied to the satellite-collision fractions obtained from the single-side neutron event sample.

Due to the dead time of the ZDC detector electronics, the timing information could only be recorded for a fraction of the triggered events. The  
size of the sample available for the analysis of time spectra does not allow for a statistically significant determination of satellite-collision fractions for each bunch pair and separation step. Therefore, one can only estimate a bunch-averaged satellite contribution. In order to improve the accuracy of the satellite estimation, the fit procedure is therefore extended with a joint likelihood maximisation, based on both timing and trigger data, at each time bin. Let $S_i$ be the number of events identified as main--satellite collisions in $T_i$ recorded events (and $t_i$ the number of trigger counts in $n_i$ sampled orbits, as defined above), the joint binomial likelihood can be written as
\begin{equation} \label{eq:localLH}
\ln\mathcal{L}_i=t_{i}\ln P_i+\left(n_{i}-t_{i}\right)\ln\left(1-P_i\right)+S_{i}\ln\left(\frac{p_{\rm{s,}\it{i}}}{P_i}\right)+(T_{i}-S_{i})\ln\left(\frac{P_i-p_{\rm{s,}\it{i}}}{P_i}\right).
\end{equation}
The maximisation procedure determines the most probable value for $p_{\rm{s},\it{i}}$ for the measured values of $n_i$, $t_i$, $T_i$ and $S_i$ and the current expected $P_i$. The $p_{\rm{s},\it{i}}$ value obtained is then fed into the global likelihood according to Eqs.~\ref{eq:likelihood} and~\ref{eq:mu}. 

In summary, the free parameters of the global likelihood fit for a given colliding bunch pair are the visible cross section, the mean values and standard deviations of the Gaussian cores of the $f(\Delta x)$ and $g(\Delta y)$ functions, and 7 to 13 tail parameters for each of the two functions.

\begin{figure}
\includegraphics[width=0.5\columnwidth]{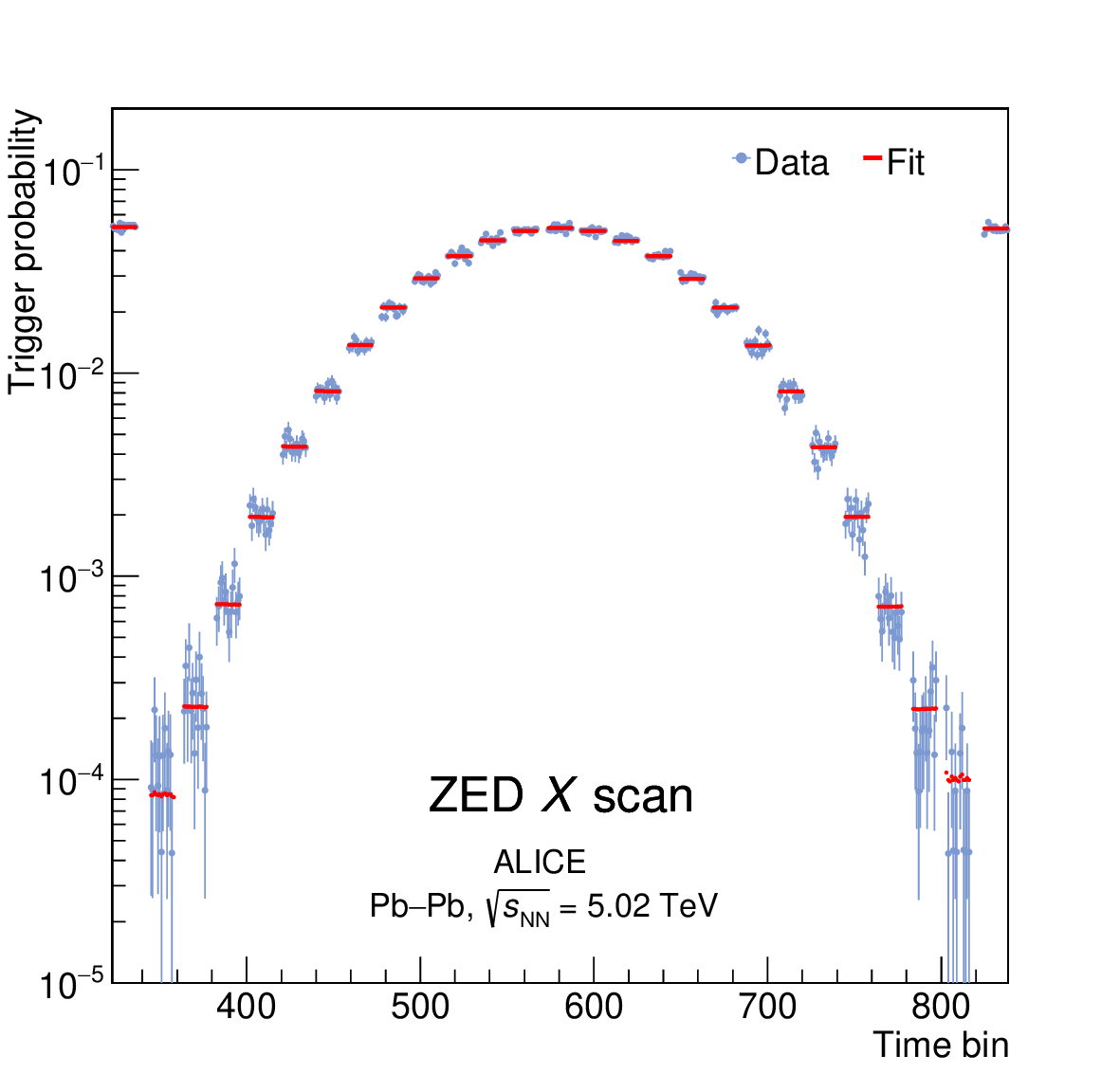}
\includegraphics[width=0.5\columnwidth]{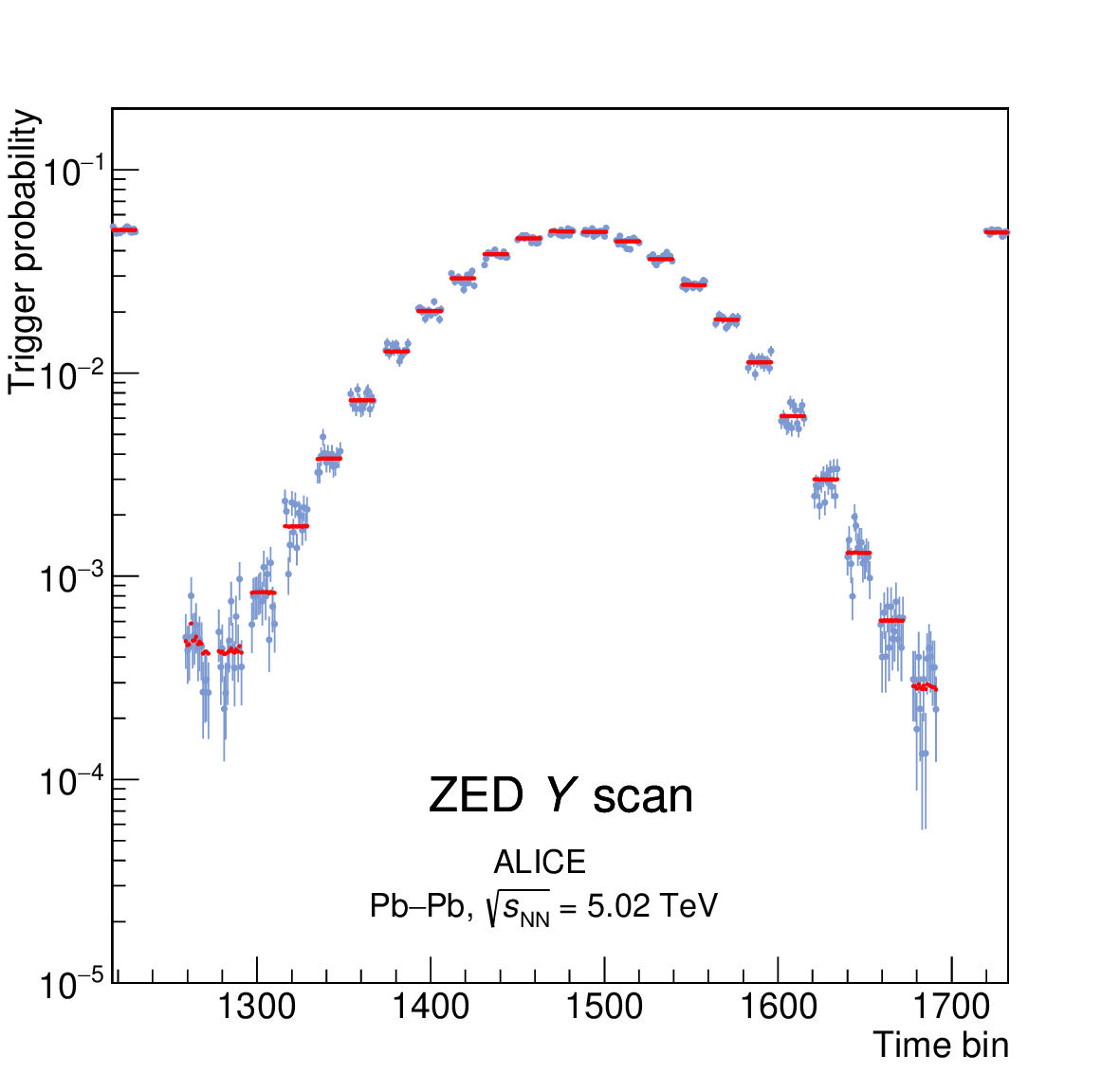}\\
\includegraphics[width=0.5\columnwidth]{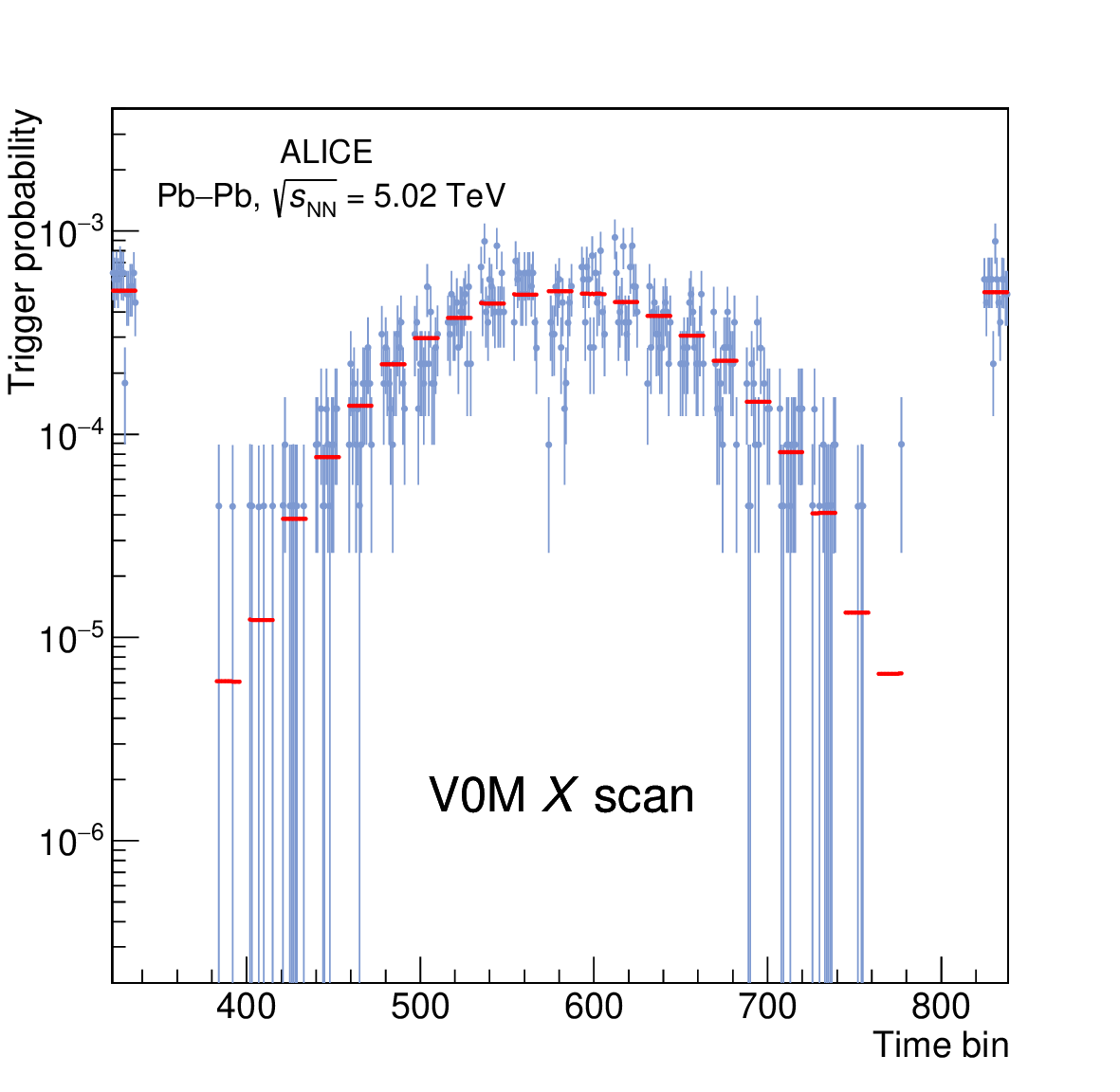}
\includegraphics[width=0.5\columnwidth]{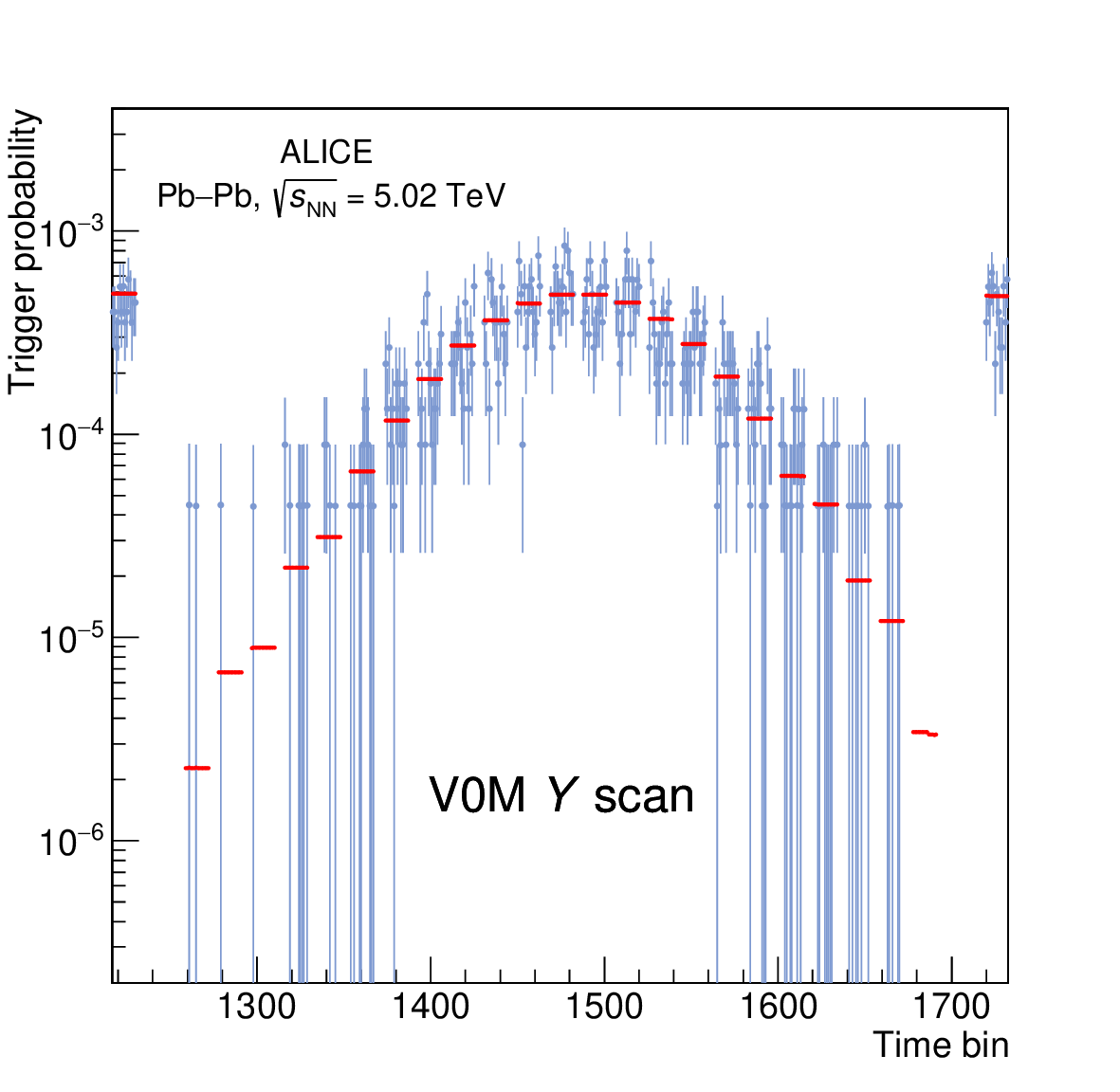}

\caption{\label{fig:probability}(Colour online) ZED (top) and V0M (bottom) trigger probabilities per bunch crossing for a typical colliding bunch pair,
as a function of time, during the first horizontal and vertical vdM scan. Each time bin corresponds to an acquisition window of $\sim 2$~s. 
The uncertainties are statistical only. The fit expectation values are also shown, as lines, in each time bin. Time bins during which the beams are being displaced, not considered in the analysis, are not shown.}

\end{figure}

As an example, in Fig.~\ref{fig:probability} the measured trigger probability per bunch crossing as a function of time during the vdM scan is shown for one pair of colliding bunches, together with the expectation from the fit.
The values of $\chi^2/\textit{ndf}$ are typically close to unity. As a remark, $\chi^2/\textit{ndf}$ values as large as $\sim 2$ are obtained if a pure Gaussian function is used, without introducing any tail parameter.

\begin{figure}
\includegraphics[width=0.5\columnwidth]{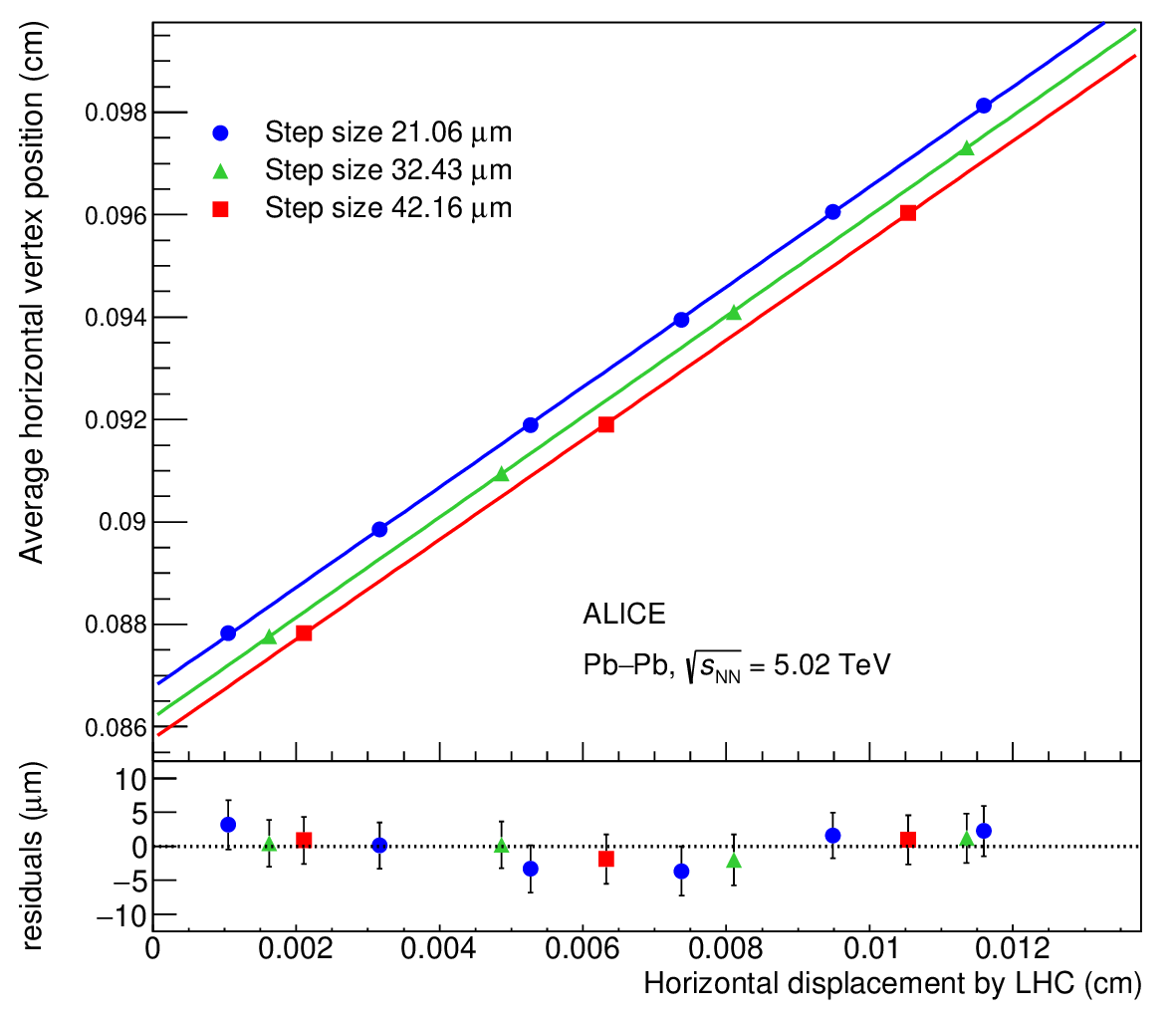}
\includegraphics[width=0.5\columnwidth]{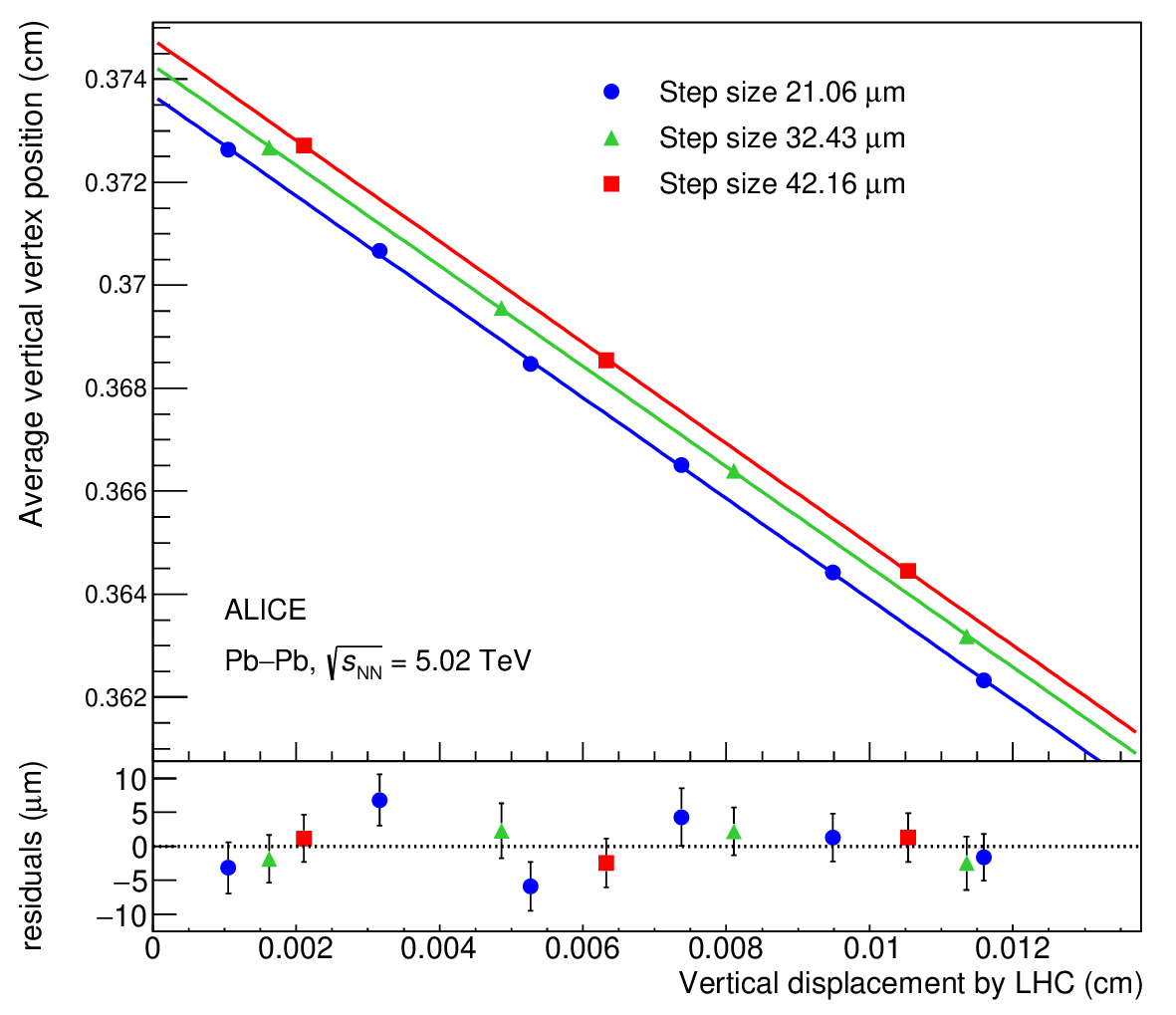}\\
\caption{\label{fig:LSC}(Colour online) Nominal versus measured displacements in the horizontal (left) and vertical (right) length-scale calibration scans, obtained from events with reconstructed-track multiplicity ranging from 260 to 500. Data are represented by symbols, while a linear fit is represented by the solid lines. The uncertainties are smaller than the symbol sizes. The fit residuals are shown in the lower panel. The blue (green, red) lines and solid circles (triangles, squares) correspond to a nominal displacement step size of 21.06 $\mu$m (32.43 $\mu$m, 42.16 $\mu$m).} 
\end{figure}

The ZED and V0M analyses provide largely independent estimates of the effective beam widths $h_{x0}$ and $h_{y0}$, via the fitted parameters of $f(\Delta x$) and $g(\Delta y)$. The $h_{x0} h_{y0}$ products obtained in the ZED and in the V0M analysis are consistent within 0.13\%, which provides an indication that detector-dependent effects such as background and pile-up are under control.

Three  length-scale calibration scans were performed for each direction, with different displacement step size, in order to test for a possible dependence on such a parameter. The horizontal (vertical) calibration factor is the slope parameter of a linear fit to the measured horizontal (vertical) vertex displacement versus the nominal one, as illustrated  in Fig.~\ref{fig:LSC}. The vertex position was determined using tracks reconstructed in the ITS. The resulting (multiplicative) correction factor to the fitted $\sigma_{\rm{vis}}$ is the product of the horizontal and vertical calibration factors, and was found to be 0.964$\pm$0.010. The uncertainty has a statistical (0.5\%) and a systematic 
contribution. The latter accounts for deviations from the linear trend in the individual fits (0.3\%), for the dependence of the results on the displacement step size (0.4\%), and for the dependence of the results on the track and event selection criteria used in the vertex determination procedure (0.7\%).

The impact of non-factorisation effects was evaluated by simultaneously fitting the rates and the luminous-region parameters (positions, sizes, transverse tilt) during both the standard and the diagonal scans with a three-dimensional non-factorisable double-Gaussian model~\cite{atlasLumi2,webbThesis,atlasLumi13TeV,aliceLumi13TeV}, and computing the bias on the head-on luminosity with respect to a factorisable model.
The resulting (multiplicative) correction factor to the fitted $\sigma_{\rm{vis}}$ is 1.011$\pm$0.011, where, conservatively, an uncertainty as large as the correction is assigned, to account for the non-accurate description of some of the luminous-region parameters by the model.

\begin{figure}
\includegraphics[width=0.5\columnwidth,viewport=0bp 0bp 566bp 348bp,clip]{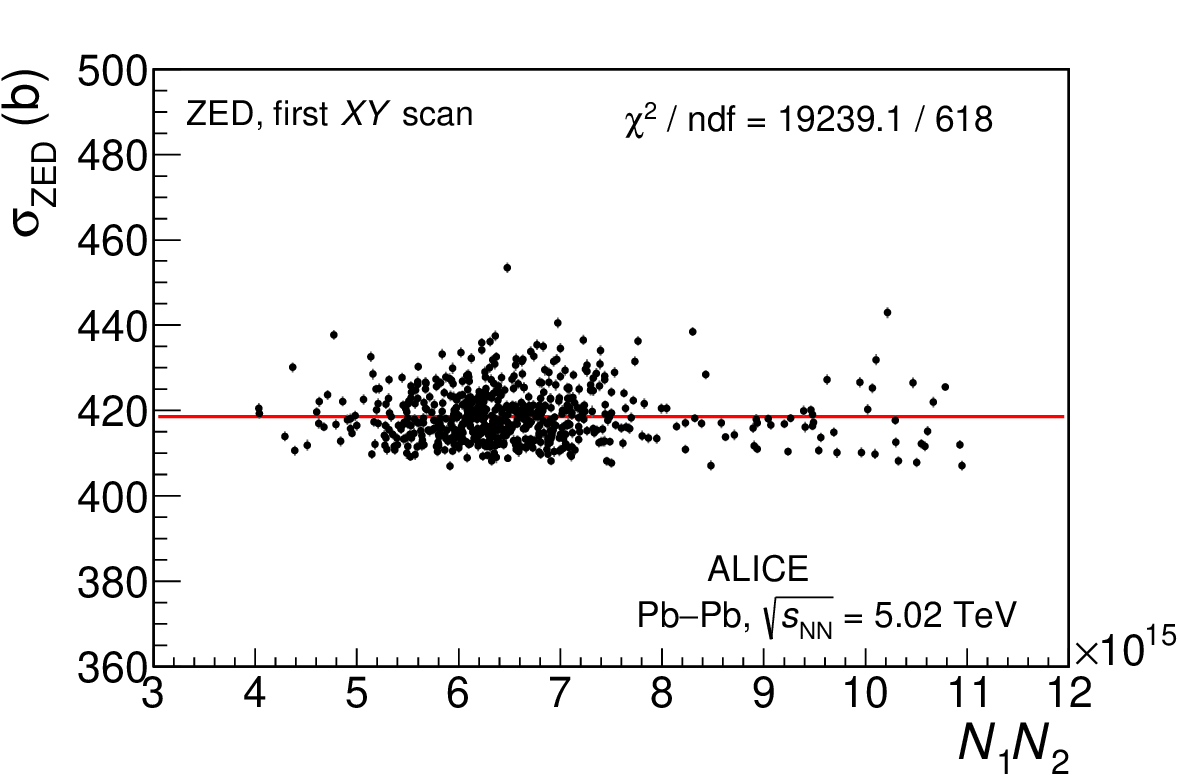}
\includegraphics[width=0.5\columnwidth,viewport=0bp 0bp 566bp 348bp,clip]{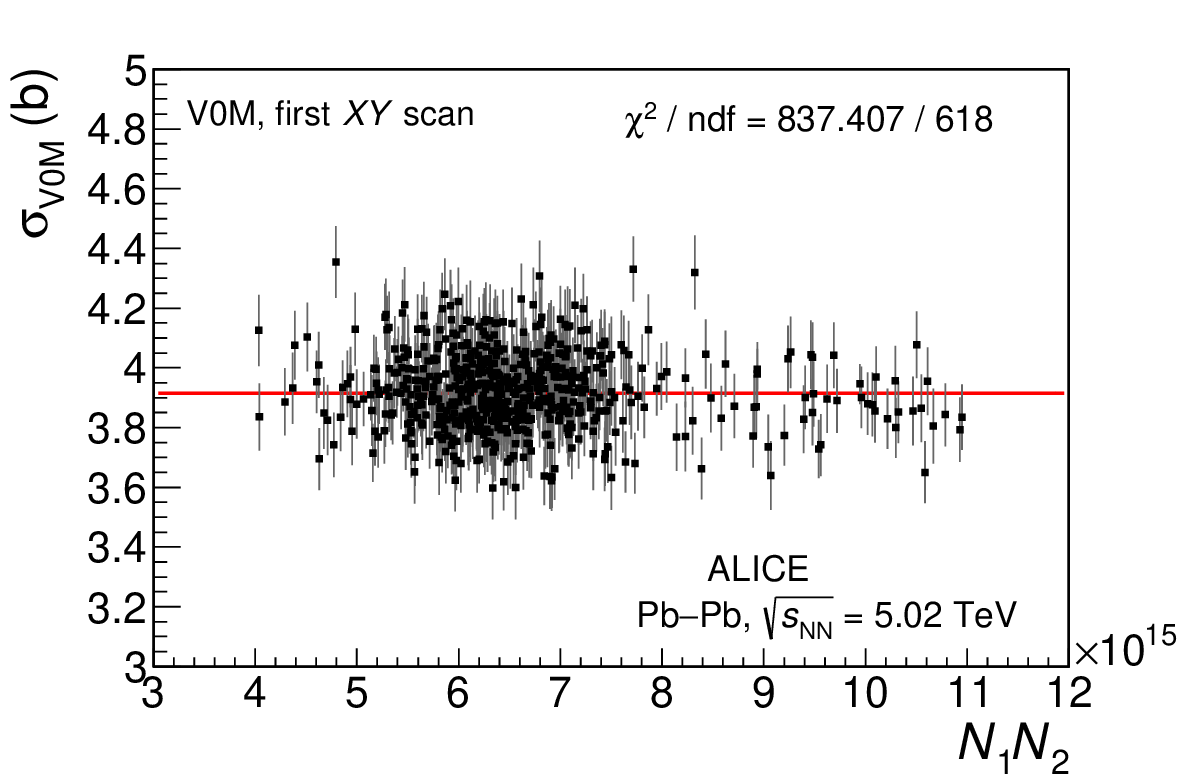}\\
\includegraphics[width=0.5\columnwidth,viewport=0bp 0bp 566bp 348bp,clip]{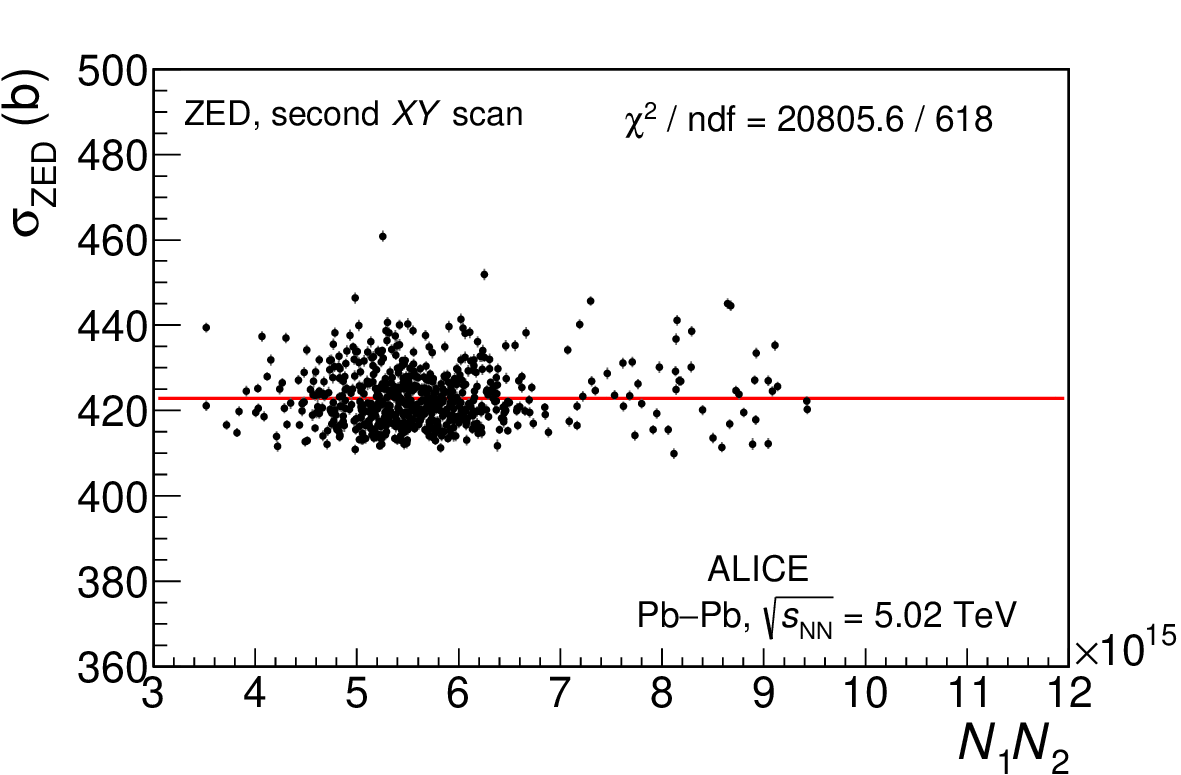}
\includegraphics[width=0.5\columnwidth,viewport=0bp 0bp 566bp 348bp,clip]{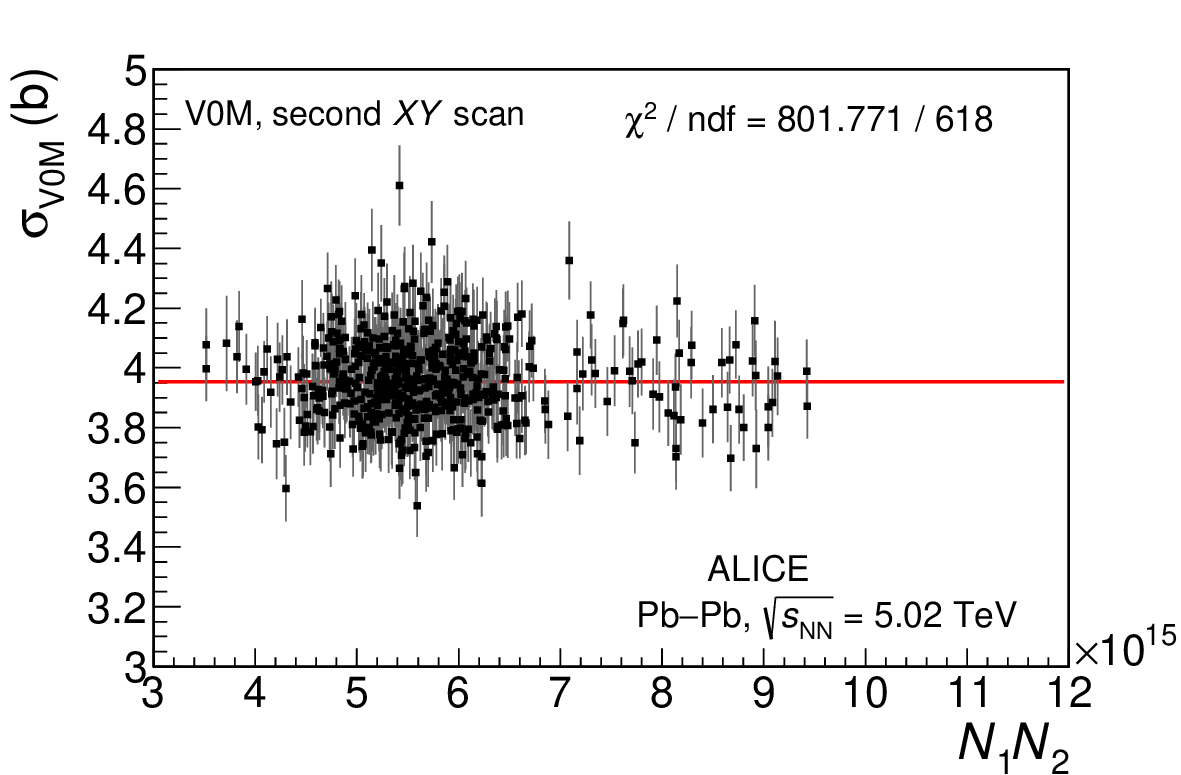}

\caption{\label{fig:sigmaVis_N1N2}(Colour online) Measured ZED (left) and V0M (right) visible cross sections as a function of the product of the ion bunch intensities, for the first (top) and second vdM scan (bottom). Uncertainties are statistical only. The solid line represents a fit to a constant value.}
\end{figure}

The ZED and V0M cross sections measured for all colliding bunch pairs and scans are shown as a function of the product of bunch intensities $N_1 N_2$ in Fig.~\ref{fig:sigmaVis_N1N2}. For both luminometers and scans, no significant dependence of $\sigma_{\rm{vis}}$ on $N_1 N_2$ is observed. However, non-statistical fluctuations of the cross section are present, particularly visible for ZED, which has better statistical precision. In order to take these into account, a systematic uncertainty of 0.1\% is assigned, computed as $\sqrt{\chi^2/\textit{ndf}-1}$ times the statistical uncertainty of the average cross section~\cite{PDG2020}, where $\chi^2/\textit{ndf}$ is obtained from the constant-value fits to the bunch-by-bunch cross sections shown in Fig.~\ref{fig:sigmaVis_N1N2}. 
The observed fluctuations are likely related to the  significant bunch-by-bunch variation of the satellite-charge fraction (on the order of 50\% root mean square, as measured by the LDM). A major contribution to the bunch-by-bunch spread of $\sigma_{\rm{vis}}$ was found to originate from pairs with large satellite-charge fraction. A bunch-by-bunch correction for satellite-charge was not performed in this analysis, due to a limited knowledge of the sensitivity of fBCT (or BPTX) to charge in satellite buckets. Instead, the bunch-averaged satellite charge was used as an overall correction to the total beam current measured by DCCT, assuming satellite charge does not contribute to the fBCT signal.

The bunch-averaged cross sections measured in the two scans agree within 1\%, which is considered as an additional systematic uncertainty. The measured visible cross sections, obtained by averaging the results from the two scans, are $\sigma_{\rm{ZED}}=$~420.58~$\pm$~0.03~(stat.)~b and $\sigma_{\rm{V0M}}=$~3.933~$\pm$~0.003~(stat.)~b. 

The combined impact of the subtraction of background from beam--gas collisions, electronic noise, and satellite collisions on the final cross section is about  1.5\% for ZED and 1\% for V0M, largely dominated by satellite collisions. The main source of uncertainty of the satellite-collision background estimation is the usage of bunch-integrated timing data in the evaluation of the satellite collision fractions, with a (potentially limited) sensitivity to bunch-by-bunch variations provided by the joint likelihood minimisation of Eq.~\ref{eq:localLH}. An alternative method was tested, where the satellite-collision probability $p_{s,i}$ for a given bunch pair is evaluated as the bunch-integrated satellite-collision fraction $S_i/T_i$ measured with the ZDC timing, scaled by the ratio between the satellite charge fraction for that bunch pair and the bunch-averaged satellite-charge fraction, both measured by the LDM. The systematic uncertainty is estimated as the maximum difference, across scans and luminometers, between the visible cross sections obtained with the standard and alternative method, and amounts to 1.2\%. The systematic uncertainty on the subtraction of background from beam--gas and electronic noise is estimated by setting the parameters $p_0$, $\tilde{p}_{1}$ and $\tilde{p}_{2}$ to zero in the likelihood fit (see Eq.~\ref{eq:likelihood} and~\ref{eq:mu}). This corresponds to the extreme assumption that all counts in nominally non-colliding bunch slots originate from collisions involving ghost charge. The variation in visible cross section, retained as uncertainty, is 0.3\% at most. 

The uncertainty of the bunch intensity is 0.8\%, from the quadratic sum of three components: 0.5\%, from the uncertainty of the total beam current normalisation from the DCCT, evaluated as described in~\cite{dcct_note}; 0.2\%, from the uncertainty of the relative bunch populations, evaluated as the difference between the fBCT- and BPTX-based results; and 0.6\%, from the uncertainty of the ghost and satellite charge~\cite{lhcbLumi2, ldm_note}, dominated by the  difference between the LHCb and LDM measurements of the ghost-charge fraction. 
No additional uncertainty is assigned to the bunch-by-bunch spread of satellite charge, because about 95\% of the bunches in each beam were colliding in IP2. Under these circumstances, the bunch-pair-averaged visible cross section is essentially driven by the total beam current measurement from DCCT (to which the sum of fBCT signals is normalised), and a non-perfect evaluation of the satellite charge in each bunch slot only leads to a bunch-by-bunch spread of the measured cross sections, with negligible or no bias to the final result, as was verified by making different assumptions for the fBCT sensitivity to satellites.

The measurement of the width of the beam-overlap region  in a van der Meer scan can be perturbed by a variation of the bunch emittance during the scan itself. The variation rate of the effective beam widths was estimated with two different methods. The first uses the difference of the measured widths between the first and second scan, the second uses the time evolution of the rate at zero separation, corrected by the bunch intensity decay. The second method yields larger variation rates (by about 70\%) than the first. The potential bias on the measured visible cross sections was estimated in a realistic simulation of the performed scans, assuming an exponential time dependence of the effective beam widths, using the slopes obtained with the second method. The resulting uncertainty is 0.5\%.  

Possible non-linearities in the steering magnet behaviour during the scan, e.g. due to hysteresis, were considered as a source of systematic uncertainty. A preliminary hysteresis model~\cite{LCMWG}
developed for the LHC was used. The model provides, for each scan step and for both beams, an upper limit to the hysteresis-induced shift of the beam position with respect to its nominal value. For this fill, the maximum shift is about 0.5~$\mu$m. In order to estimate a possible bias on the cross section, the fit of Eq.~\ref{eq:likelihood} was performed with the separation at each step modified according to the predicted position shift of both beams. The change in average visible cross section is 0.2\% for both luminometers and is retained as a systematic uncertainty.

The uncertainty of the orbit-drift correction was conservatively taken to be as large as the effect of the correction (0.15\%). 
The uncertainty of the beam--beam deflection correction was evaluated by varying the input parameters to the deflection calculation within a reasonable range, as described in~\cite{aliceLumi13TeV}, and found to be less than 0.1\%. 
The effect of distortions of the bunch shapes due to the mutual interaction between the two beams was also evaluated, within the framework outlined in~\cite{balaguraBeamBeam}, and found to be less than 0.1\%.

The systematic uncertainty associated with the choice of the fitting strategy was evaluated: by varying the range of beam separations described by the Gaussian core (varying thereby the number of fit parameters used to describe the tails); by discarding the last scan step, where the satellite contribution is dominant; and by extracting the visible cross section from a simultaneous fit to all colliding bunch pairs, with common shape parameters, instead of averaging the results from individual fits. The resulting uncertainty is 0.4\%.

The total systematic uncertainty of the visible cross section measurement, obtained as the quadratic sum of the contributions listed above, amounts to 2.4\% for both ZED and V0M.

\subsection{Hadronic inelastic cross section determination}

\begin{figure}[!ht]
    \centering
    \includegraphics[width=0.49\columnwidth]{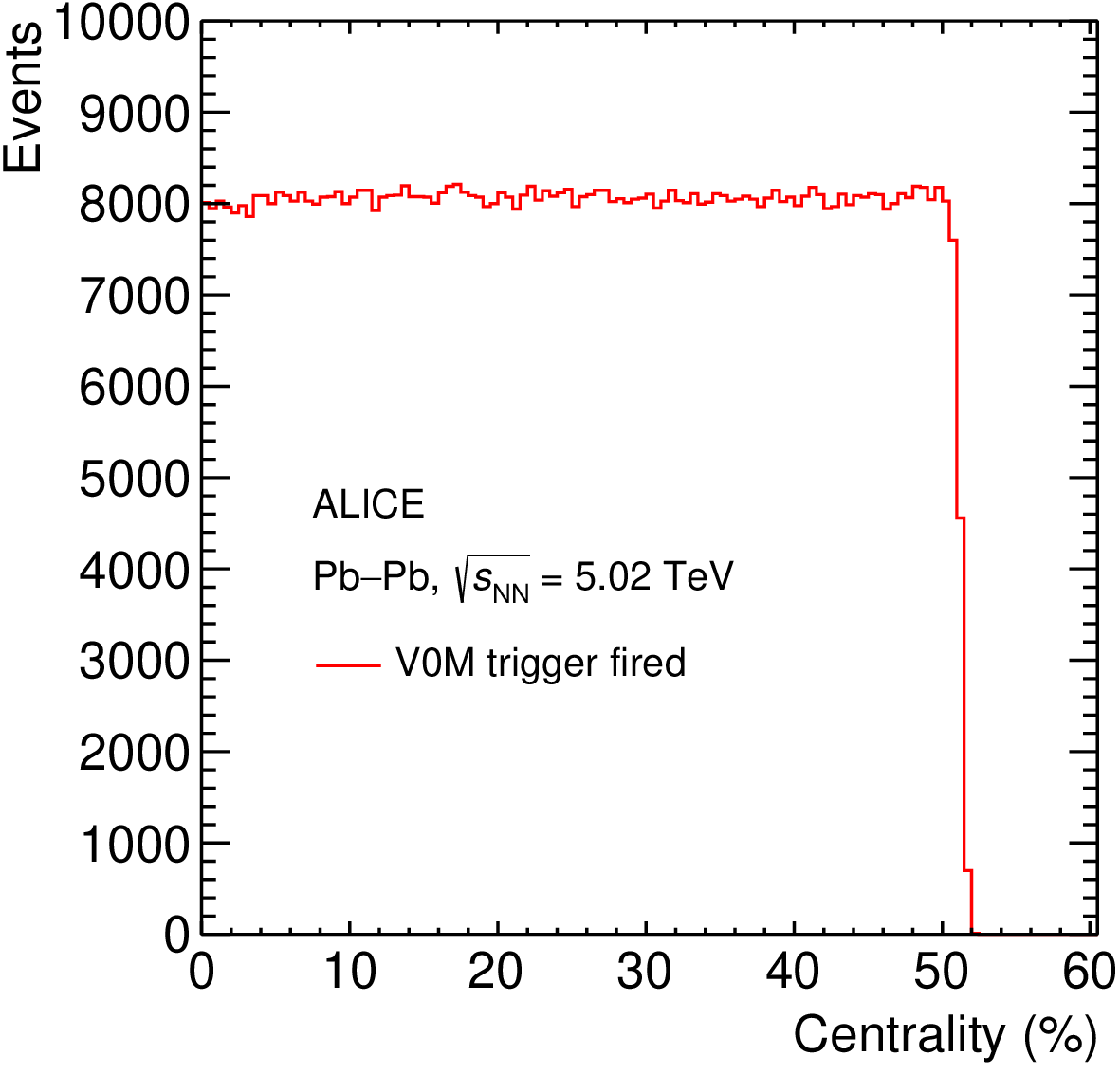}
    \caption{\label{fig:V0Mefficiency} Centrality distribution of events satisfying the V0M trigger condition, for the LHC fill during which the vdM scan was performed.
    }
\end{figure}

As an additional output of the vdM scan analysis, the inelastic hadronic cross section $\sigma_{\rm{had}}$ was determined by correcting the visible cross section $\sigma_{\rm{V0M}}$ for the V0M trigger efficiency. The ALICE centrality determination framework~\cite{aliCent1, aliCent2} assigns to each event a centrality value, based on the total signal amplitude in the V0 detector. The centrality is defined as the probability that a hadronic Pb--Pb collision results in an amplitude larger than the measured value. The centrality calibration for the 2018 sample was performed using a mimimum-bias trigger requiring a signal in each of V0A, V0C, ZNA and ZNC. Such a trigger is fully efficient for hadronic events and free from electromagnetic contamination for the $\sim$~90\% most central events~\cite{aliMultPbPbPaper, aliMultPbPbPN}. In order to obtain the shape of the amplitude spectrum in the most peripheral events, the minimum-bias-triggered spectrum is fitted with a Monte Carlo implementation of the Glauber model~\cite{Glauber}, coupled with a two-component ancestor model for particle production; the fit is performed above a chosen amplitude threshold (anchor point, corresponding to a centrality of 90\%), where no trigger bias is expected. The centrality distribution of V0M-triggered events, determined using the framework described above, is shown in Fig.~\ref{fig:V0Mefficiency}. The distribution is uniform in the 0--50\% centrality range, where the V0M trigger is fully efficient, then drops rapidly to zero in the range 50--52\%. When the distribution is normalised such that its integral in 0--50\% is 0.5, its total integral provides the V0M efficiency for hadronic interactions, $\epsilon_{\rm{had}}$. For the fill in which the vdM scan was performed, this procedure results in $\epsilon_{\rm{had}}$~=~0.513~$\pm$~0.012. The quoted uncertainty is systematic and is obtained as the quadratic sum of two components. The first one, of 1.4\%, was determined, similar to what was done in Ref.~\cite{aliMultPbPbPaper}, by varying the centrality at the anchor point within $\pm$~1\% (referring here to an absolute variation, i.e., from 89\% to 91\%). The second one, of 1.8\%, was determined as the difference between the default efficiency value and the one obtained by fitting the V0 amplitude spectrum with a different template, based on the T$_{\rm R}$ENTo model~\cite{trento}.   Finally, one has
\begin{displaymath}
    \sigma_{\rm{had}} = \frac{\sigma_{\rm{V0M}}}{\epsilon_{\rm{had}}} = 7.67 \pm 0.25\;\rm{~b},
\end{displaymath}
where the quoted uncertainty is the combination of the statistical and systematic uncertainties of the visible cross section, of 2.4\%, and of the trigger efficiency, of 2.3\%. The measured cross section is in agreement with the  prediction of (7.62~$\pm$~0.15)~b from Ref.~\cite{Glauber2}, based on a Monte Carlo implementation of the Glauber model with a nuclear radius of $\sim$~6.7~fm, a nuclear skin depth for protons (neutrons) of $\sim$~0.45~fm ($\sim$~0.56~fm),  and an inelastic nucleon--nucleon cross section of $\sim$~67~mb. 

\subsection{Consistency and stability of the luminosity calibration}

In order to test the stability and mutual consistency of the ZED and V0M calibrations, the luminosities measured with the two reference signals throughout the whole 2015 and 2018 data-taking periods were compared on a run-by-run basis. In the ALICE nomenclature, a run is a set of data collected within a start and a stop of the data acquisition, under stable detector and trigger configurations~\footnote{For the data-taking period under consideration, the duration of a run ranges from $\sim$~5~minutes to $\sim$~7~hours.}. For each run, the trigger counts, integrated over colliding bunch slots, were corrected by subtracting the estimated beam--gas background, detector noise, and background from main--satellite collisions. As explained earlier, the beam--gas background was estimated by means of the counts in non-colliding bunch slots, rescaled by the relative fractions of beam intensities; the contribution from detector noise was estimated via the counts in empty slots; the background from main--satellite collisions was estimated using the ZDC timing data. For each run, the pile-up corrected ratio between the V0M- and \hbox{ZED-based} luminosities was computed from the corrected number of trigger counts $N_{\rm{V0M}}$ and $N_{\rm{ZED}}$ and from the total number of bunch crossings in the run $N_{\rm{BC}}$ as \begin{equation}\frac{L_{\rm{V0M}}}{L_{\rm{ZED}}} = \frac{\ln{(1-N_{\rm{V0M}}/N_{\rm{BC}})}\sigma_{\rm{ZED}}}{\ln{(1-N_{\rm{ZED}}/N_{\rm{BC}})}\sigma_{\rm{V0M}}}.
\label{eq:V0MtoZED}
\end{equation}

 While the ZED trigger settings remained unchanged throughout the 2015 and 2018 data-taking periods, the threshold for the V0M trigger was different in 2015 and 2018. Furthermore, in 2018, the threshold was slightly adjusted a few times during data-taking as the V0M-based centrality trigger was being tuned. For the data-taking periods with different threshold settings with respect to the vdM scan, the V0M trigger efficiency was re-determined with the procedure described earlier, and the V0M cross section re-scaled by the ratio of the measured efficiency to that measured in the fill containing the van der Meer scans.

\begin{figure}
    \includegraphics[width=0.49\columnwidth]{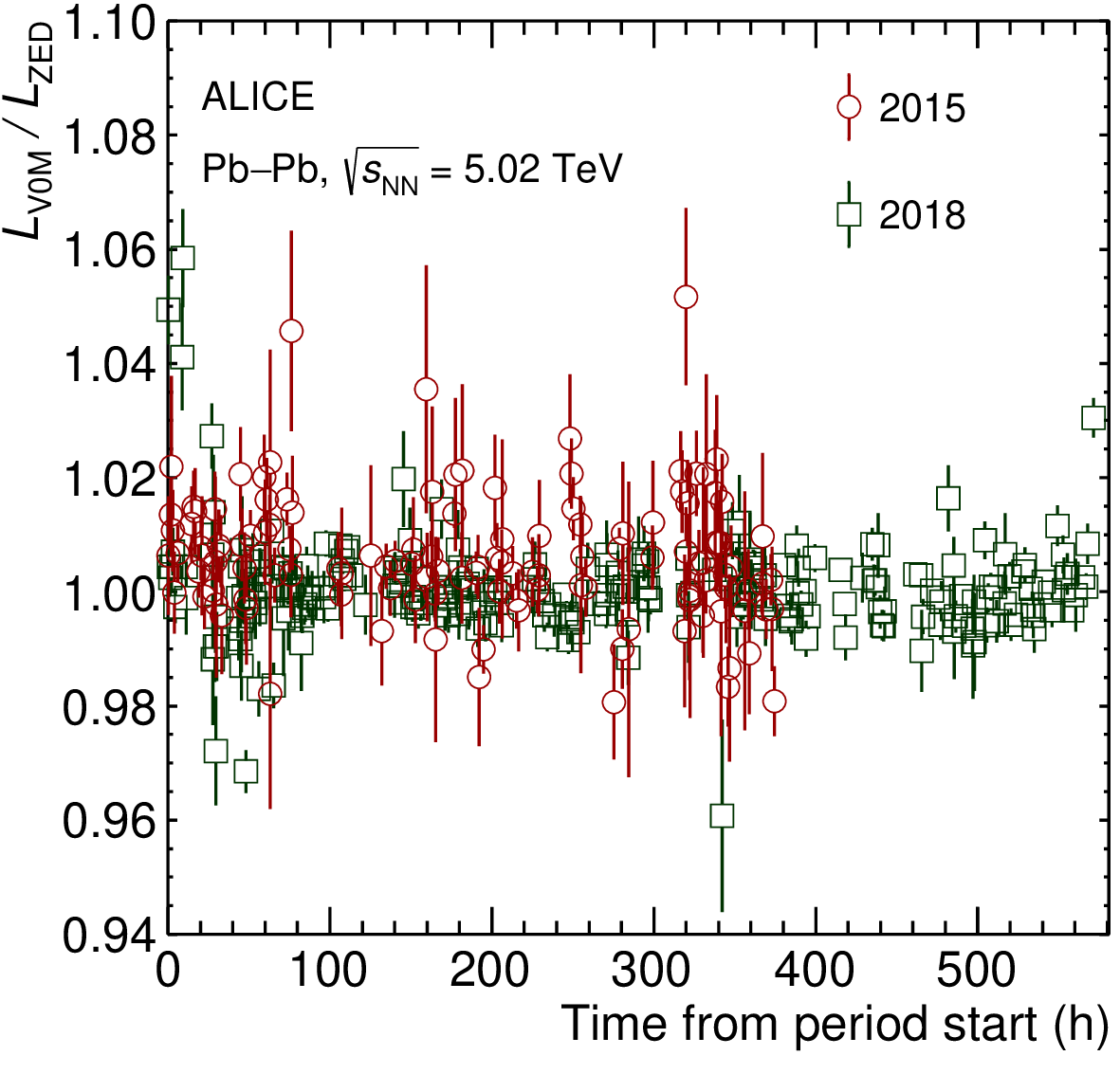}
    \includegraphics[width=0.49\columnwidth]{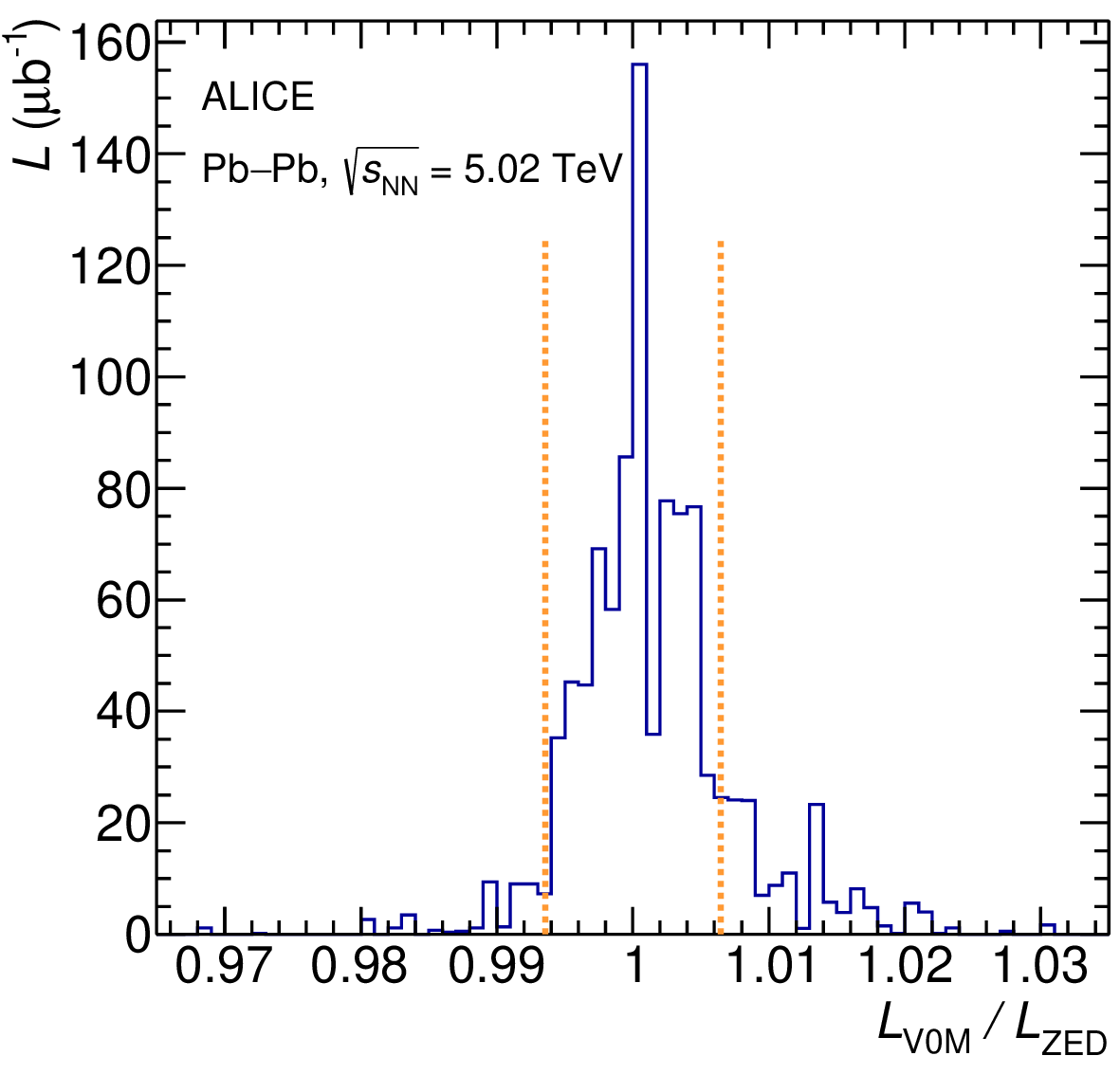}
    \caption{\label{fig:stability}(Colour online) Left: ratio of V0M and ZED  luminosities calculated according to Eq.~\ref{eq:V0MtoZED} as a function of time  with respect to the beginning of the data-taking periods for 2015 (circles) and 2018 (squares). The uncertainties are statistical. Right: distribution of the ratios for all runs (2015 and 2018), weighted with the run luminosity. The dashed vertical lines are located at $L_{\rm{V0M}}/L_{\rm{ZED}}=1$~$\pm$~$\alpha$, where $\alpha$~$\sim$~0.007 is the mean quadratic difference from unity.}
\end{figure}

The luminosity ratio as a function of time and the distribution of the ratio values over all runs, weighted with the run luminosity, are shown in Fig.~\ref{fig:stability}. The mean quadratic difference of the ratio from unity is about 0.7\% and is retained as a systematic uncertainty of the stability and mutual consistency of the luminosity calibration. When the analysis is restricted to the 2015 or 2018 sample, the mean quadratic difference from unity amounts to 1\% or 0.5\%, respectively.

\subsection{Luminosity uncertainty}

In Table~\ref{tab:uncertainties} a summary of the different contributions to the uncertainty of the visible cross section and the luminosity measurement is presented. The luminosity uncertainty, obtained as the quadratic sum of the visible cross section uncertainty and of the stability and consistency uncertainty, amounts to 2.5\% for both ZED and V0M. For the sake of comparison, the luminosity uncertainty obtained by ALICE  for Pb--Pb collisions at \snn~$=$~2.76~TeV (LHC Run~1) was of 5--6\%~\cite{aliPerf}.

\begin{table}[tbp]
\caption{Relative uncertainties of the measurement of visible cross sections and luminosity in Pb--Pb collisions at \fivenn. The stability and consistency and the total luminosity uncertainties refer to the full Run 2 sample (2015 and 2018); uncertainties for the single periods are given in the text.}
  \begin{center}
    \begin{tabular}{l|c} 
      Source & Uncertainty (\%) \\ 
      &ZED $|$ V0M \\ \hline
      Statistical & \hspace{0.5 mm}0.008 $|$ 0.08 \\
      $h_{x0}h_{y0}$ consistency (V0M vs ZED) &  0.13 \\
      Length-scale calibration &  1 \\
      Non-factorisation & 1.1 \\
      Bunch-to-bunch consistency & 0.1  \\
      Scan-to-scan consistency & 1 \\
      Satellite collisions & 1.2  \\
      Beam--gas and noise & 0.3  \\
      Bunch intensity & 0.8 \\ 
      Emittance variation & 0.5 \\
      Magnetic non-linearities & 0.2 \\ 
      Orbit drift & 0.15 \\
      Beam--beam deflection and distortion & 0.1 \\
      Fitting scheme & 0.4 \\ \hline
      Total of visible cross section & 2.4 \\ \hline
      Stability and consistency & 0.7  \\ \hline
      Total of luminosity & 2.5 $|$ 2.5 \\ \hline
    \end{tabular}
  \end{center}
  \vspace{0.2 cm}
  \label{tab:uncertainties}
\end{table}

\section{Conclusions
\label{sec:Conclusions}}
In 2015 and 2018, the ALICE Collaboration took data with \PbPb collisions at a centre-of-mass energy \fivenn. In order to provide a reference for the luminosity determination, vdM scans were performed and visible cross 
sections were measured for two processes, ZED (neutron emission in the acceptance of the neutron Zero Degree Calorimeters) and V0M (energy deposition in the V0 detector by events up to $\sim$~50\% centrality). Each of the two detectors provides a measurement of the luminosity with a total uncertainty, for the full sample (2015 and 2018), of 2.5\%. These uncertainties improve by about a factor of two with respect to those obtained by ALICE in previous studies dedicated to Pb--Pb collisions at \snn~$=$~2.76 TeV.  
The inelastic cross section for hadronic interactions in \PbPb collisions at \fivenn, obtained by trigger-efficiency correction of the V0M cross section, was measured to be 7.67~$\pm$~0.25~b, in agreement with predictions from the Glauber model.


\newenvironment{acknowledgement}{\relax}{\relax}
\begin{acknowledgement}
\section*{Acknowledgements}

The ALICE Collaboration would like to thank all its engineers and technicians for their invaluable contributions to the construction of the experiment and the CERN accelerator teams for the outstanding performance of the LHC complex.
The ALICE Collaboration gratefully acknowledges the resources and support provided by all Grid centres and the Worldwide LHC Computing Grid (WLCG) collaboration.
The ALICE Collaboration acknowledges the following funding agencies for their support in building and running the ALICE detector:
A. I. Alikhanyan National Science Laboratory (Yerevan Physics Institute) Foundation (ANSL), State Committee of Science and World Federation of Scientists (WFS), Armenia;
Austrian Academy of Sciences, Austrian Science Fund (FWF): [M 2467-N36] and Nationalstiftung f\"{u}r Forschung, Technologie und Entwicklung, Austria;
Ministry of Communications and High Technologies, National Nuclear Research Center, Azerbaijan;
Conselho Nacional de Desenvolvimento Cient\'{\i}fico e Tecnol\'{o}gico (CNPq), Financiadora de Estudos e Projetos (Finep), Funda\c{c}\~{a}o de Amparo \`{a} Pesquisa do Estado de S\~{a}o Paulo (FAPESP) and Universidade Federal do Rio Grande do Sul (UFRGS), Brazil;
Bulgarian Ministry of Education and Science, within the National Roadmap for Research Infrastructures 2020-2027 (object CERN), Bulgaria;
Ministry of Education of China (MOEC) , Ministry of Science \& Technology of China (MSTC) and National Natural Science Foundation of China (NSFC), China;
Ministry of Science and Education and Croatian Science Foundation, Croatia;
Centro de Aplicaciones Tecnol\'{o}gicas y Desarrollo Nuclear (CEADEN), Cubaenerg\'{\i}a, Cuba;
Ministry of Education, Youth and Sports of the Czech Republic, Czech Republic;
The Danish Council for Independent Research | Natural Sciences, the VILLUM FONDEN and Danish National Research Foundation (DNRF), Denmark;
Helsinki Institute of Physics (HIP), Finland;
Commissariat \`{a} l'Energie Atomique (CEA) and Institut National de Physique Nucl\'{e}aire et de Physique des Particules (IN2P3) and Centre National de la Recherche Scientifique (CNRS), France;
Bundesministerium f\"{u}r Bildung und Forschung (BMBF) and GSI Helmholtzzentrum f\"{u}r Schwerionenforschung GmbH, Germany;
General Secretariat for Research and Technology, Ministry of Education, Research and Religions, Greece;
National Research, Development and Innovation Office, Hungary;
Department of Atomic Energy Government of India (DAE), Department of Science and Technology, Government of India (DST), University Grants Commission, Government of India (UGC) and Council of Scientific and Industrial Research (CSIR), India;
National Research and Innovation Agency - BRIN, Indonesia;
Istituto Nazionale di Fisica Nucleare (INFN), Italy;
Japanese Ministry of Education, Culture, Sports, Science and Technology (MEXT) and Japan Society for the Promotion of Science (JSPS) KAKENHI, Japan;
Consejo Nacional de Ciencia (CONACYT) y Tecnolog\'{i}a, through Fondo de Cooperaci\'{o}n Internacional en Ciencia y Tecnolog\'{i}a (FONCICYT) and Direcci\'{o}n General de Asuntos del Personal Academico (DGAPA), Mexico;
Nederlandse Organisatie voor Wetenschappelijk Onderzoek (NWO), Netherlands;
The Research Council of Norway, Norway;
Commission on Science and Technology for Sustainable Development in the South (COMSATS), Pakistan;
Pontificia Universidad Cat\'{o}lica del Per\'{u}, Peru;
Ministry of Education and Science, National Science Centre and WUT ID-UB, Poland;
Korea Institute of Science and Technology Information and National Research Foundation of Korea (NRF), Republic of Korea;
Ministry of Education and Scientific Research, Institute of Atomic Physics, Ministry of Research and Innovation and Institute of Atomic Physics and University Politehnica of Bucharest, Romania;
Ministry of Education, Science, Research and Sport of the Slovak Republic, Slovakia;
National Research Foundation of South Africa, South Africa;
Swedish Research Council (VR) and Knut \& Alice Wallenberg Foundation (KAW), Sweden;
European Organization for Nuclear Research, Switzerland;
Suranaree University of Technology (SUT), National Science and Technology Development Agency (NSTDA), Thailand Science Research and Innovation (TSRI) and National Science, Research and Innovation Fund (NSRF), Thailand;
Turkish Energy, Nuclear and Mineral Research Agency (TENMAK), Turkey;
National Academy of  Sciences of Ukraine, Ukraine;
Science and Technology Facilities Council (STFC), United Kingdom;
National Science Foundation of the United States of America (NSF) and United States Department of Energy, Office of Nuclear Physics (DOE NP), United States of America.
In addition, individual groups or members have received support from:
Marie Sk\l{}odowska Curie, Strong 2020 - Horizon 2020, European Research Council (grant nos. 824093, 896850, 950692), European Union;
Academy of Finland (Center of Excellence in Quark Matter) (grant nos. 346327, 346328), Finland;
Programa de Apoyos para la Superaci\'{o}n del Personal Acad\'{e}mico, UNAM, Mexico.

\end{acknowledgement} 

\bibliographystyle{utphys}   
\bibliography{bibliography}

\newpage
\appendix

%


\section{Fitting function definition}
\label{app:function}

The luminosity dependence on the horizontal separation is parametrised (see Eq.~\ref{eq:R_vis}) with the fitting function $f(\Delta x)$.

With 25 scan steps, one can choose $-12$~$\leq$~$j$~$\leq$~$+12$, so that the nominal separation at step $j$ is given by \begin{equation} \Delta x_{{\rm{nom,}}j} = \frac{j}{24}(\Delta x_{\rm{nom,max}}-\Delta x_{\rm{nom,min}}),
\end{equation}
where  $\Delta x_{\rm{nom,max}}$~=~$-\Delta x_{\rm{nom,min}}$~=~97.3~$\mu$m, and $j$~=~0 denotes the (nominal) zero separation. As discussed in Section~\ref{sec:results}, the actual separation $\Delta x_j$ is obtained by correcting the nominal separation $\Delta x_{{\rm{nom,}}j}$ for the orbit drift and beam--beam deflection effects.

With the above convention, the fitting function is defined as
\begin{equation}\label{eq:fitFunc}
  f(\Delta x_j) = e^{-\frac{(\Delta x_j - \mu + \delta_j)^2}{2\sigma^2}},
\end{equation}
with
\begin{equation}
    \delta_j = \delta_{-j}
\end{equation}
and
\begin{equation}
  \delta_j = 0\;\;{\rm{for}}\;\;|j| < j_0,
\end{equation}
where $j_0$~$>$~0 is the threshold chosen for the transition between the Gaussian core and the tail (see Section~\ref{sec:results} for details). The fit parameters in the function are the mean value $\mu$, the standard deviation $\sigma$ and the offsets $\delta_j$ (with $j_0$~$<$~$j$~$\leq$~12). 

The definition of the function $g(\Delta y)$ used to parametrise the luminosity dependence on the vertical separation is identical, with independent offset parameters.

Depending on the considered colliding bunch pair, the fitting functions use 6$~\leq~j_0~\leq~$9.

\newpage

\section{The ALICE Collaboration}
\label{app:collab}
\begin{flushleft} 
\small

S.~Acharya\,\orcidlink{0000-0002-9213-5329}\,$^{\rm 123,131}$, 
D.~Adamov\'{a}\,\orcidlink{0000-0002-0504-7428}\,$^{\rm 85}$, 
A.~Adler$^{\rm 69}$, 
G.~Aglieri Rinella\,\orcidlink{0000-0002-9611-3696}\,$^{\rm 32}$, 
M.~Agnello\,\orcidlink{0000-0002-0760-5075}\,$^{\rm 29}$, 
N.~Agrawal\,\orcidlink{0000-0003-0348-9836}\,$^{\rm 50}$, 
Z.~Ahammed\,\orcidlink{0000-0001-5241-7412}\,$^{\rm 131}$, 
S.~Ahmad\,\orcidlink{0000-0003-0497-5705}\,$^{\rm 15}$, 
S.U.~Ahn\,\orcidlink{0000-0001-8847-489X}\,$^{\rm 70}$, 
I.~Ahuja\,\orcidlink{0000-0002-4417-1392}\,$^{\rm 37}$
% ...

\section*{Affiliation Notes}

$^{\rm I}$ Deceased\\
$^{\rm II}$ Also at: Max-Planck-Institut f\"{u}r Physik, Munich, Germany\\
$^{\rm III}$ Also at: Italian National Agency for New Technologies, Energy and Sustainable Economic Development (ENEA), Bologna, Italy\\
$^{\rm IV}$ Also at: Dipartimento DET del Politecnico di Torino, Turin, Italy\\
$^{\rm V}$ Also at: Department of Applied Physics, Aligarh Muslim University, Aligarh, India\\
$^{\rm VI}$ Also at: Institute of Theoretical Physics, University of Wroclaw, Poland\\
$^{\rm VII}$ Also at: An institution covered by a cooperation agreement with CERN\\

\section*{Collaboration Institutes}

$^{1}$ A.I. Alikhanyan National Science Laboratory (Yerevan Physics Institute) Foundation, Yerevan, Armenia\\
$^{2}$ AGH University of Krakow, Cracow, Poland\\
$^{3}$ Bogolyubov Institute for Theoretical Physics, National Academy of Sciences of Ukraine, Kiev, Ukraine\\
$^{4}$ Bose Institute, Department of Physics  and Centre for Astroparticle Physics and Space Science (CAPSS), Kolkata, India\\
$^{5}$ California Polytechnic State University, San Luis Obispo, California, United States\\
$^{6}$ Central China Normal University, Wuhan, China\\
$^{7}$ Centro de Aplicaciones Tecnol\'{o}gicas y Desarrollo Nuclear (CEADEN), Havana, Cuba\\
$^{8}$ Centro de Investigaci\'{o}n y de Estudios Avanzados (CINVESTAV), Mexico City and M\'{e}rida, Mexico\\
$^{9}$ Chicago State University, Chicago, Illinois, United States\\
$^{10}$ China Institute of Atomic Energy, Beijing, China\\
$^{11}$ Chungbuk National University, Cheongju, Republic of Korea\\
$^{12}$ Comenius University Bratislava, Faculty of Mathematics, Physics and Informatics, Bratislava, Slovak Republic\\
$^{13}$ COMSATS University Islamabad, Islamabad, Pakistan\\
$^{14}$ Creighton University, Omaha, Nebraska, United States\\
$^{15}$ Department of Physics, Aligarh Muslim University, Aligarh, India\\
$^{16}$ Department of Physics, Pusan National University, Pusan, Republic of Korea\\
$^{17}$ Department of Physics, Sejong University, Seoul, Republic of Korea\\
$^{18}$ Department of Physics, University of California, Berkeley, California, United States\\
$^{19}$ Department of Physics, University of Oslo, Oslo, Norway\\
$^{20}$ Department of Physics and Technology, University of Bergen, Bergen, Norway\\
$^{21}$ Dipartimento di Fisica, Universit\`{a} di Pavia, Pavia, Italy\\
$^{22}$ Dipartimento di Fisica dell'Universit\`{a} and Sezione INFN, Cagliari, Italy\\
$^{23}$ Dipartimento di Fisica dell'Universit\`{a} and Sezione INFN, Trieste, Italy\\
$^{24}$ Dipartimento di Fisica dell'Universit\`{a} and Sezione INFN, Turin, Italy\\
$^{25}$ Dipartimento di Fisica e Astronomia dell'Universit\`{a} and Sezione INFN, Bologna, Italy\\
$^{26}$ Dipartimento di Fisica e Astronomia dell'Universit\`{a} and Sezione INFN, Catania, Italy\\
$^{27}$ Dipartimento di Fisica e Astronomia dell'Universit\`{a} and Sezione INFN, Padova, Italy\\
$^{28}$ Dipartimento di Fisica `E.R.~Caianiello' dell'Universit\`{a} and Gruppo Collegato INFN, Salerno, Italy\\
$^{29}$ Dipartimento DISAT del Politecnico and Sezione INFN, Turin, Italy\\
$^{30}$ Dipartimento di Scienze MIFT, Universit\`{a} di Messina, Messina, Italy\\
$^{31}$ Dipartimento Interateneo di Fisica `M.~Merlin' and Sezione INFN, Bari, Italy\\
$^{32}$ European Organization for Nuclear Research (CERN), Geneva, Switzerland\\
$^{33}$ Faculty of Electrical Engineering, Mechanical Engineering and Naval Architecture, University of Split, Split, Croatia\\
$^{34}$ Faculty of Engineering and Science, Western Norway University of Applied Sciences, Bergen, Norway\\
$^{35}$ Faculty of Nuclear Sciences and Physical Engineering, Czech Technical University in Prague, Prague, Czech Republic\\
$^{36}$ Faculty of Physics, Sofia University, Sofia, Bulgaria\\
$^{37}$ Faculty of Science, P.J.~\v{S}af\'{a}rik University, Ko\v{s}ice, Slovak Republic\\
$^{38}$ Frankfurt Institute for Advanced Studies, Johann Wolfgang Goethe-Universit\"{a}t Frankfurt, Frankfurt, Germany\\
$^{39}$ Fudan University, Shanghai, China\\
$^{40}$ Gangneung-Wonju National University, Gangneung, Republic of Korea\\
$^{41}$ Gauhati University, Department of Physics, Guwahati, India\\
$^{42}$ Helmholtz-Institut f\"{u}r Strahlen- und Kernphysik, Rheinische Friedrich-Wilhelms-Universit\"{a}t Bonn, Bonn, Germany\\
$^{43}$ Helsinki Institute of Physics (HIP), Helsinki, Finland\\
$^{44}$ High Energy Physics Group,  Universidad Aut\'{o}noma de Puebla, Puebla, Mexico\\
$^{45}$ Horia Hulubei National Institute of Physics and Nuclear Engineering, Bucharest, Romania\\
$^{46}$ Indian Institute of Technology Bombay (IIT), Mumbai, India\\
$^{47}$ Indian Institute of Technology Indore, Indore, India\\
$^{48}$ INFN, Laboratori Nazionali di Frascati, Frascati, Italy\\
$^{49}$ INFN, Sezione di Bari, Bari, Italy\\
$^{50}$ INFN, Sezione di Bologna, Bologna, Italy\\
$^{51}$ INFN, Sezione di Cagliari, Cagliari, Italy\\
$^{52}$ INFN, Sezione di Catania, Catania, Italy\\
$^{53}$ INFN, Sezione di Padova, Padova, Italy\\
$^{54}$ INFN, Sezione di Pavia, Pavia, Italy\\
$^{55}$ INFN, Sezione di Torino, Turin, Italy\\
$^{56}$ INFN, Sezione di Trieste, Trieste, Italy\\
$^{57}$ Inha University, Incheon, Republic of Korea\\
$^{58}$ Institute for Gravitational and Subatomic Physics (GRASP), Utrecht University/Nikhef, Utrecht, Netherlands\\
$^{59}$ Institute of Experimental Physics, Slovak Academy of Sciences, Ko\v{s}ice, Slovak Republic\\
$^{60}$ Institute of Physics, Homi Bhabha National Institute, Bhubaneswar, India\\
$^{61}$ Institute of Physics of the Czech Academy of Sciences, Prague, Czech Republic\\
$^{62}$ Institute of Space Science (ISS), Bucharest, Romania\\
$^{63}$ Institut f\"{u}r Kernphysik, Johann Wolfgang Goethe-Universit\"{a}t Frankfurt, Frankfurt, Germany\\
$^{64}$ Instituto de Ciencias Nucleares, Universidad Nacional Aut\'{o}noma de M\'{e}xico, Mexico City, Mexico\\
$^{65}$ Instituto de F\'{i}sica, Universidade Federal do Rio Grande do Sul (UFRGS), Porto Alegre, Brazil\\
$^{66}$ Instituto de F\'{\i}sica, Universidad Nacional Aut\'{o}noma de M\'{e}xico, Mexico City, Mexico\\
$^{67}$ iThemba LABS, National Research Foundation, Somerset West, South Africa\\
$^{68}$ Jeonbuk National University, Jeonju, Republic of Korea\\
$^{69}$ Johann-Wolfgang-Goethe Universit\"{a}t Frankfurt Institut f\"{u}r Informatik, Fachbereich Informatik und Mathematik, Frankfurt, Germany\\
$^{70}$ Korea Institute of Science and Technology Information, Daejeon, Republic of Korea\\
$^{71}$ KTO Karatay University, Konya, Turkey\\
$^{72}$ Laboratoire de Physique Subatomique et de Cosmologie, Universit\'{e} Grenoble-Alpes, CNRS-IN2P3, Grenoble, France\\
$^{73}$ Lawrence Berkeley National Laboratory, Berkeley, California, United States\\
$^{74}$ Lund University Department of Physics, Division of Particle Physics, Lund, Sweden\\
$^{75}$ Nagasaki Institute of Applied Science, Nagasaki, Japan\\
$^{76}$ Nara Women{'}s University (NWU), Nara, Japan\\
$^{77}$ National and Kapodistrian University of Athens, School of Science, Department of Physics , Athens, Greece\\
$^{78}$ National Centre for Nuclear Research, Warsaw, Poland\\
$^{79}$ National Institute of Science Education and Research, Homi Bhabha National Institute, Jatni, India\\
$^{80}$ National Nuclear Research Center, Baku, Azerbaijan\\
$^{81}$ National Research and Innovation Agency - BRIN, Jakarta, Indonesia\\
$^{82}$ Niels Bohr Institute, University of Copenhagen, Copenhagen, Denmark\\
$^{83}$ Nikhef, National institute for subatomic physics, Amsterdam, Netherlands\\
$^{84}$ Nuclear Physics Group, STFC Daresbury Laboratory, Daresbury, United Kingdom\\
$^{85}$ Nuclear Physics Institute of the Czech Academy of Sciences, Husinec-\v{R}e\v{z}, Czech Republic\\
$^{86}$ Oak Ridge National Laboratory, Oak Ridge, Tennessee, United States\\
$^{87}$ Ohio State University, Columbus, Ohio, United States\\
$^{88}$ Physics department, Faculty of science, University of Zagreb, Zagreb, Croatia\\
$^{89}$ Physics Department, Panjab University, Chandigarh, India\\
$^{90}$ Physics Department, University of Jammu, Jammu, India\\
$^{91}$ Physics Department, University of Rajasthan, Jaipur, India\\
$^{92}$ Physics Program and International Institute for Sustainability with Knotted Chiral Meta Matter (SKCM2), Hiroshima University, Hiroshima, Japan\\
$^{93}$ Physikalisches Institut, Eberhard-Karls-Universit\"{a}t T\"{u}bingen, T\"{u}bingen, Germany\\
$^{94}$ Physikalisches Institut, Ruprecht-Karls-Universit\"{a}t Heidelberg, Heidelberg, Germany\\
$^{95}$ Physik Department, Technische Universit\"{a}t M\"{u}nchen, Munich, Germany\\
$^{96}$ Politecnico di Bari and Sezione INFN, Bari, Italy\\
$^{97}$ Research Division and ExtreMe Matter Institute EMMI, GSI Helmholtzzentrum f\"ur Schwerionenforschung GmbH, Darmstadt, Germany\\
$^{98}$ Saha Institute of Nuclear Physics, Homi Bhabha National Institute, Kolkata, India\\
$^{99}$ School of Physics and Astronomy, University of Birmingham, Birmingham, United Kingdom\\
$^{100}$ Secci\'{o}n F\'{\i}sica, Departamento de Ciencias, Pontificia Universidad Cat\'{o}lica del Per\'{u}, Lima, Peru\\
$^{101}$ Stefan Meyer Institut f\"{u}r Subatomare Physik (SMI), Vienna, Austria\\
$^{102}$ SUBATECH, IMT Atlantique, Nantes Universit\'{e}, CNRS-IN2P3, Nantes, France\\
$^{103}$ Suranaree University of Technology, Nakhon Ratchasima, Thailand\\
$^{104}$ Technical University of Ko\v{s}ice, Ko\v{s}ice, Slovak Republic\\
$^{105}$ The Henryk Niewodniczanski Institute of Nuclear Physics, Polish Academy of Sciences, Cracow, Poland\\
$^{106}$ The University of Texas at Austin, Austin, Texas, United States\\
$^{107}$ Universidad Aut\'{o}noma de Sinaloa, Culiac\'{a}n, Mexico\\
$^{108}$ Universidade de S\~{a}o Paulo (USP), S\~{a}o Paulo, Brazil\\
$^{109}$ Universidade Estadual de Campinas (UNICAMP), Campinas, Brazil\\
$^{110}$ Universidade Federal do ABC, Santo Andre, Brazil\\
$^{111}$ University of Cape Town, Cape Town, South Africa\\
$^{112}$ University of Houston, Houston, Texas, United States\\
$^{113}$ University of Jyv\"{a}skyl\"{a}, Jyv\"{a}skyl\"{a}, Finland\\
$^{114}$ University of Kansas, Lawrence, Kansas, United States\\
$^{115}$ University of Liverpool, Liverpool, United Kingdom\\
$^{116}$ University of Science and Technology of China, Hefei, China\\
$^{117}$ University of South-Eastern Norway, Kongsberg, Norway\\
$^{118}$ University of Tennessee, Knoxville, Tennessee, United States\\
$^{119}$ University of the Witwatersrand, Johannesburg, South Africa\\
$^{120}$ University of Tokyo, Tokyo, Japan\\
$^{121}$ University of Tsukuba, Tsukuba, Japan\\
$^{122}$ University Politehnica of Bucharest, Bucharest, Romania\\
$^{123}$ Universit\'{e} Clermont Auvergne, CNRS/IN2P3, LPC, Clermont-Ferrand, France\\
$^{124}$ Universit\'{e} de Lyon, CNRS/IN2P3, Institut de Physique des 2 Infinis de Lyon, Lyon, France\\
$^{125}$ Universit\'{e} de Strasbourg, CNRS, IPHC UMR 7178, F-67000 Strasbourg, France, Strasbourg, France\\
$^{126}$ Universit\'{e} Paris-Saclay, Centre d'Etudes de Saclay (CEA), IRFU, D\'{e}partment de Physique Nucl\'{e}aire (DPhN), Saclay, France\\
$^{127}$ Universit\'{e}  Paris-Saclay, CNRS/IN2P3, IJCLab, Orsay, France\\
$^{128}$ Universit\`{a} degli Studi di Foggia, Foggia, Italy\\
$^{129}$ Universit\`{a} del Piemonte Orientale, Vercelli, Italy\\
$^{130}$ Universit\`{a} di Brescia, Brescia, Italy\\
$^{131}$ Variable Energy Cyclotron Centre, Homi Bhabha National Institute, Kolkata, India\\
$^{132}$ Warsaw University of Technology, Warsaw, Poland\\
$^{133}$ Wayne State University, Detroit, Michigan, United States\\
$^{134}$ Westf\"{a}lische Wilhelms-Universit\"{a}t M\"{u}nster, Institut f\"{u}r Kernphysik, M\"{u}nster, Germany\\
$^{135}$ Wigner Research Centre for Physics, Budapest, Hungary\\
$^{136}$ Yale University, New Haven, Connecticut, United States\\
$^{137}$ Yonsei University, Seoul, Republic of Korea\\
$^{138}$  Zentrum  f\"{u}r Technologie und Transfer (ZTT), Worms, Germany\\
$^{139}$ Affiliated with an institute covered by a cooperation agreement with CERN\\
$^{140}$ Affiliated with an international laboratory covered by a cooperation agreement with CERN.\\

\end{flushleft} 

\end{document}